\documentclass[12pt,preprint]{aastex}

\newcommand{\lya}{Ly$\alpha$}
\newcommand{\lyb}{Ly$\beta$}

\newcommand{\hi}{H~{\sc i}}
\newcommand{\oi}{O~{\sc i}}
\newcommand{\cii}{C~{\sc ii}}
\newcommand{\civ}{C~{\sc iv}}
\newcommand{\siii}{Si~{\sc ii}}
\newcommand{\siiv}{Si~{\sc iv}}

\newcommand{\mgii}{Mg~{\sc ii}}
\newcommand{\feii}{Fe~{\sc ii}}
\newcommand{\ovi}{O~{\sc vi}}
\newcommand{\mhi}{{\rm H \; \mbox{\tiny I}}}

\newcommand{\kms}{km~s$^{-1}$}

\newcommand{\zq}{z_{\rm QSO}}
\newcommand{\chisq}{$\chi^2$}
\newcommand{\rchisq}{$\chi^2_{\rm r}$}

\newcommand{\taueffa}{$\tau_{\rm eff}^{\alpha}$}
\newcommand{\taueffb}{$\tau_{\rm eff}^{\beta}$}
\newcommand{\G}{$\Gamma_{-12}$}

\shorttitle{Evolution of optical depth in the \lya\ forest} 

\shortauthors{Becker et al.}

\begin{document}

\title{The Evolution of Optical Depth in the \lya\ Forest:\\ 
       Evidence Against Reionization at $z \sim 6$\altaffilmark{1}}

\author{George D. Becker\altaffilmark{2}, Michael
        Rauch\altaffilmark{3}, Wallace L. W. Sargent\altaffilmark{2}}

\altaffiltext{1}{The observations were made at the W.M. Keck
  Observatory which is operated as a scientific partnership between
  the California Institute of Technology and the University of
  California; it was made possible by the generous support of the
  W.M. Keck Foundation.}
\altaffiltext{2}{Palomar Observatory, California Institute of
  Technology, Pasadena, CA 91125, USA; gdb@astro.caltech.edu,
  wws@astro.caltech.edu}
\altaffiltext{3}{Carnegie Observatories, 813 Santa Barbara Street,
  Pasadena, CA 91101, USA; mr@ociw.edu}

\begin{abstract}

  We examine the evolution of the IGM \lya\ optical depth distribution
  using the transmitted flux probability distribution function (PDF)
  in a sample of 63 QSOs spanning absorption redshifts $1.7 < z <
  5.8$.  The data are compared to two theoretical $\tau$
  distributions: a model distribution based on the density
  distribution of \citet{me00} (MHR00), and a lognormal distribution.
  We assume a uniform UV background and an isothermal IGM for the
  MHR00 model, as has been done in previous works where transmitted
  flux statistics have been used to infer an abrupt change in the IGM
  at $z \sim 6$.  Under these assumptions, the MHR00 model produces
  poor fits to the observed flux PDFs at redshifts where the optical
  depth distribution is well sampled, unless large continuum
  corrections are applied.  However, the lognormal $\tau$ distribution
  fits the data at all redshifts with only minor continuum
  adjustments.  We use a simple parametrization for the evolution of
  the lognormal parameters to calculate the expected mean transmitted
  flux at $z > 5.4$.  The lognormal $\tau$ distribution predicts the
  observed \lya\ and \lyb\ effective optical depths at $z > 5.7$ while
  simultaneously fitting the mean transmitted flux down to $z = 1.6$.
  In contrast, the best-fitting power-law under-predicts the amount of
  absorption both at $z > 5.7$ and at $z < 2.5$.  If the evolution of
  the lognormal distribution at $z < 5$ reflects a slowly-evolving
  density field, temperature, and UV background, then no sudden change
  in the IGM at $z \sim 6$ due to late reionization appears necessary.
  We have used the lognormal optical depth distribution without any
  assumption about the underlying density field.  If the MHR00 density
  distribution is correct, then a non-uniform UV background and/or IGM
  temperature may be required to produce the correct flux PDF.  We
  find that an inverse temperature-density relation greatly improves
  the PDF fits, but with a large scatter in the equation of state
  index.  The lognormal $\tau$ distribution therefore currently offers
  the best match to the observed flux PDF and the most reliable
  predictor for the transmitted flux at high redshift.
  
\end{abstract}

\keywords{cosmology: observations --- cosmology: early universe ---
  intergalactic medium --- quasars: absorption lines}

\section{Introduction}\label{sec:fluxpdf_intro}

The \lya\ forest serves as our most fundamental probe of the evolution
of the intergalactic medium (IGM).  While numerous models have been
proposed for the underlying density field (see Rauch~1998 for a
review), the current consensus is a self-gravitating network of
filamentary structures collapsing out of initially Gaussian density
perturbations.  Given a description of the IGM that relates density
and transmitted flux, one can compute various cosmological parameters
and examine the large-scale evolution of the Universe.

Perhaps the most dramatic inferences drawn from the evolution of \lya\
transmitted flux is that the reionization of the IGM may have ended as
late as $z \sim 6.2$ \citep{becker01,white03,fan02,fan06}.  This
conclusion is based not only on the appearance of complete
Gunn-Peterson troughs in the spectra of QSOs at $z > 6$, but on the
accelerated decline and increased variance in the mean transmitted
flux at $z > 5.7$ \citep{fan06}.  Late reionization is potentially at
odds with the transmitted flux seen towards the highest-redshift known
QSO, SDSS~J1148$+$5251 \citep[$z_{\rm QSO} =
6.42$,][]{white03,white05,ohfur05}.  In addition, the fact that the
observed number density of \lya-emitting galaxies does not evolve
strongly from $z \sim 5.7$ to $z \sim 6.5$ implies that the IGM is
already highly ionized at these redshifts
\citep{hu04,hu06,mr04,mr06,stern05}.  Galactic winds \citep{santos04}
or locally ionized bubbles
\citep{haiman05,wyithe05,furlanetto04,furlanetto06} may allow \lya\
photons to escape even if the IGM is significantly neutral.
Additional arguments may be made about the thermal history of the IGM
\citep{theuns02,hui03} or the apparent size of the transmission
regions around $z \sim 6$ QSOs
\citep{mesinger04a,mesinger04b,wyithe04,fan06}.  However, the
evolution of the \lya\ forest remains the strongest evidence for late
reionization.

Still, the significance of the disappearance of transmitted flux at $z
\sim 6$ has been highly debated \citep{songaila02,songaila04,lidz06a}.
As \citet{songaila02} pointed out, the mean transmitted flux in an
inhomogeneous IGM will depend strongly on the underlying density
distribution, or more precisely, on the optical depth distribution.
At $z \sim 6$, any transmitted flux will arise from rare voids, which
lie in the tail of the optical depth distribution.  Using a sample of
19 QSOs at $z > 5.7$, \citet{fan06} showed that the evolution the mean
transmitted flux at $z \sim 6$ diverges significantly from that
expected for a commonly-used model of the IGM density
\citep[][referred to herein as MHR00]{me00}.  The question, then, is
whether the MHR00 model describes the distribution of optical depths
accurately enough to make reliable predictions at very high redshift.

In this paper we examine two theoretical optical depth distributions
and their predictions for the \lya\ transmitted flux probability
distribution function (PDF) The first is based on the gas density
distribution given by MHR00, which has been used to make claims of
late reionization.  Their density distribution is derived from simple
arguments about the dynamics of the IGM (see
\S\ref{sec:theoretical_pdfs}) and matches the output of an earlier
numerical simulation \citep{me96}.  In order to compute optical
depths, assumptions must be made about the ionizing background and the
thermal state of the IGM.  As other authors have done, we will
primarily consider a uniform UV background and an isothermal IGM.  In
\S\ref{sec:temp_density} we will briefly generalize to a
non-isothermal equation of state.

The second case we consider is a simple lognormal optical depth
distribution.  This choice can be motivated in at least two ways.
Initially Gaussian density perturbations will give rise to a lognormal
density field when the initial peculiar velocity field is also
Gaussian \citep{coles91}.  Indeed, \citet{bi92} demonstrated that a
lognormal density distribution can produce many properties of the
observed \lya\ forest \citep[see also][]{bi95,bi97}.  More generally,
however, a lognormal distribution naturally arises as a result of the
central limit theorem when a quantity is determined by several
multiplicative factors.  For optical depth, these factors are density,
temperature, and ionization rate.  Here we will consider the lognormal
distribution to be a generic distribution with the desirable
properties of being non-zero and having a potentially large variance.
Our main conclusions will not depend on any assumptions about the
underlying density field.

The transmitted flux PDF has been used to constrain a variety of
cosmological parameters
\citep[e.g.,][]{rauch97,gaztanaga99,mcdonald00,choudhury01,desjacques05,
  lidz06b}, with many authors assuming an optical depth distribution
similar to one we consider here.  We will examine the distributions
themselves and their evolution with redshift by attempting to fit the
models to the observed flux PDFs from a large sample of Keck HIRES
data spanning \lya\ absorption redshifts $1.7 < z < 5.8$.  We
introduce the data in \S\ref{sec:fluxpdf_data}.  In
\S\ref{sec:flux_pdfs} the optical depth distributions are derived and
used to fit the observed flux PDFs.  We find that the lognormal
distribution provides a better fit to the data at all redshifts where
the optical depth distributions are well sampled.  In
\S\ref{sec:redshift_evolution} we perform a simple fit to evolution of
the lognormal distribution and use it to predict the mean transmitted
flux at $z > 5.7$.  In \S\ref{sec:temp_density} we modify the model
distribution by applying a non-isothermal equation of state.  Finally,
our results are summarized in \S\ref{sec:fluxpdf_conclusions}.


\section{The Data}\label{sec:fluxpdf_data}

Observations were made using the HIRES spectrograph \citep{vogt94} on
Keck I between 1993 and 2006.  Targets are listed in
Table~\ref{tab:fitted_regions}.  QSOs at $\zq < 4.8$ were observed
using the original HIRES CCD and were reduced using the MAKEE package
written by Tom Barlow.  QSOs at $\zq > 4.8$ were observed using the
upgraded detector and reduced using a custom set of IDL routines as
described in \citet{becker06}.  The IDL package is based on the
optimal sky subtraction technique of \citet{kelson03}.  For nearly all
of our observations we used an 0\farcs86 slit, which gives a velocity
resolution FWHM of $\Delta v = 6.7$~\kms.

We will return to the issue of continuum fitting in
\S\ref{sec:fitting_pdfs}.  For now we will describe our baseline
fitting procedure for quasars at various redshifts.  For objects at
$\zq < 4.8$, individual exposures were typically bright enough that a
continuum could be fit to individual orders.  This was done by hand
using a slowly varying spline fit.  The orders were then normalized
prior to combining.  At higher redshifts, we performed a relative flux
calibration of each exposure using standard stars.  The individual
exposures were then combined prior to continuum fitting.  A spline fit
was again used for QSOs at $\zq \le 5.4$.  However, since the
transmission regions at $z > 5$ rarely, if ever, reach the continuum,
the fits were of a very low order and intended only to emulate the
general structure of continua observed in lower redshift QSOs
\citep[e.g.,][]{telfer02,suzuki06}.  For $\zq > 5.7$ we used a power
law fit to the continuum of the form $f_{\nu} \propto \nu^{-0.5}$.

Determining a quasar continuum is generally a subjective process whose
accuracy will depend strongly on how much of the continuum has been
absorbed (see Lidz et al.~2004b for a discussion).  At $z \sim 3$,
much of the spectrum will still be unabsorbed and errors in the
continuum fit will depend on signal-to-noise of the data and the
personal bias of the individual performing the fit.  For high-quality
data, errors in the continuum at $z \sim 3$ should be $\lesssim 1\%$.
This uncertainty will increase with redshift as more of the continuum
gets absorbed.  By $z \sim 5.5$, very few transmission regions remain
and the continuum must be inferred from the slope of the spectrum
redward of the \lya\ emission line.  However, the spectral slope may
have an unseen break near \lya.  In addition, echelle data are
notoriously difficult to accurately flux calibrate.  We therefore
expect our power-law continuum estimates at $z \sim 6$ to be off by as
much as a factor of two.

\section{Flux Probability Distribution Functions}\label{sec:flux_pdfs}

\subsection{Observed PDFs}\label{sec:observed_pdfs}

Observed transmitted flux probability distribution functions (PDFs)
were taken from spectra of the 63 quasars listed in
Table~\ref{tab:fitted_regions}.  In order to avoid contamination from
the proximity region and from \ovi/\lyb absorption, we limited our
analysis to pixels 10000~\kms\ redward of the \lya\ emission line and
5000~\kms\ redward of the \ovi\ emission line.  The offsets were made
intentionally large to account for possible errors in the QSO
redshifts.  In order for each region to contain enough pixels to be
statistically significant yet avoid strong redshift evolution within a
sample, we divided the \lya\ forest in each sightline into two
sections covering $\sim 60$~\AA\ rest wavelength.  Regions containing
damped \lya\ systems were discarded.  We further exclude wavelengths
covered by the telluric A and B bands.  Other atmospheric absorption
due to water vapor was typically weak compared to the \lya\ absorption
at the same wavelength and so was ignored.
Table~\ref{tab:fitted_regions} lists the redshift interval for each
region of the \lya\ forest we examine.

Metal lines can be a significant contaminant in the \lya\ forest,
particularly at lower redshifts.  We therefore removed as many lines
as could be identified either by damped \lya\ absorption or from
multiple metal lines at the same redshift.  In addition to the
doublets \civ, \siiv, and \mgii, we searched for coincidences of
\siii, Si~{\sc iii}, \cii, \oi, \feii, Al~{\sc ii}, and Al~{\sc iii}.
For exceptionally strong systems we also masked weaker lines such as
Cr~{\sc ii}, Ti~{\sc ii}, S~{\sc ii}, and Zn~{\sc ii}.  Lines in the
forest were masked according to the structure and extent of lines
identified redward of \lya\ emission.  Very strong line that could be
identified only from their presence in the \lya\ forest (e.g.,
saturated \civ) were also masked.  However, we did not mask weak lines
found in the forest without counterparts redward of \lya\ emission.
Doing so would preferentially discard pixels with low \lya\ optical
depth (where the metal lines can be seen), introducing a potentially
larger bias in the PDF than the one incurred by leaving the
contaminated pixels in the sample.  In any case, our primary concern
is with strong metal lines that could mimic saturated \lya\
absorption.  Weak metal lines are not expected to significantly alter
the flux PDF.

The observed transmitted flux PDF for each region was computed in
normalized flux bins of 0.02.  Errors were computed using bootstrap
resampling \citep{press92}.  Each region was divided into many short
sections spanning 200~\kms, and 1000 replicates of each region were
constructed by randomly drawing sections with replacement.  For this
work we have used only the diagonal elements of the error matrix.  As
noted by \citet{mcdonald00} and \citet{desjacques05}, ignoring the
off-diagonal elements when performing \chisq\ fitting can have a
significant effect on the width of the \chisq\ distribution, but has
only a small effect on the values of the best-fit parameters.  For
comparison, we have repeated the analyses presented in this paper
using purely Poisson errors and have obtained nearly identical
results.

\subsection{Theoretical PDFs}\label{sec:theoretical_pdfs}

We will examine two possible distributions for \lya\ optical depths:
one based on the gas density distribution given by MHR00, and the
other a lognormal distribution.  In this section we derive the
expected flux PDF for each case.

\subsubsection{MHR00 model}\label{sec:mhr00_model}

The MHR00 gas density distribution is derived analytically based on
assuming that the density fluctuations are initially Gaussian, that
the gas in voids is expanding at constant velocities, and that the
densities are smoothed on the Jeans length of the photoionized gas.
The resulting parametric form for the volume-weighted density
distribution is
\begin{equation}
   P_{\Delta}^{\rm MHR00}(\Delta) = A \exp{\left[ 
                           -\frac{(\Delta^{-2/3} - C_0)^2}{2(2\delta_0/3)^2}
                           \right]} \Delta^{-\beta} \, ,
   \label{eq:mhr00_Pdelta}
\end{equation}
where $\Delta \equiv \rho/\bar{\rho}$ is the gas overdensity and $A$,
$C_0$, $\delta_0$, and $\beta$ are constants.  We take $\delta_0 =
7.61/(1+z)$ and $\beta$ from Table~1 of MHR00, which produces good
fits to $\Lambda$CDM ($\Omega_{\rm m} = 0.4$) simulation of
\citet{me96}.  We then set $A$ and $C_0$ such that the total area
under $P_{\Delta}^{\rm MHR00}(\Delta)$ and the mean overdensity are
both equal to one.  Parameters for redshifts other than those listed
in MHR00 are linearly interpolated.

To convert from densities to optical depths, assumptions must be made
about the ionizing background radiation and the thermal state of the
gas.  The \lya\ optical depth of a uniform IGM would be
\begin{equation}
   \tau_u = \frac{\pi e^2}{m_e c} f_{\alpha} \lambda_{\alpha} H^{-1}(z)n_{\mhi} \, ,
   \label{eq:tau_u}
\end{equation}
where $f_{\alpha}$ is the \lya\ oscillator strength, $\lambda_{\alpha}
= 1216$~\AA, and $H(z)$ is the Hubble constant at redshift $z$
\citep{gp65}.  In the case of photoionization equilibrium, the optical
depth $\tau(\Delta)$ for an overdensity $\Delta$ can be expressed in
terms of the \hi\ ionization rate $\Gamma$, and the recombination
coefficient $\alpha$ as \citep{weinberg97}
\begin{equation}
   \tau(\Delta) \propto 
        \frac{(1+z)^{4.5}(\Omega_{\rm b}h^2)^2 \alpha[T(\Delta)]}
             {h \Gamma(\Delta,z) \Omega_{\rm m}^{0.5}} \Delta^2 \, ,
   \label{eq:tau_delta_propto}
\end{equation}
where $\alpha$ depends on the temperature as $\alpha(T) \propto
T^{-0.7}$ for $T \sim 10^4$~K \citep{abel97}.  The IGM temperature
will generally depend on the density, which is typically expressed as
a power-law equation of state, $T(\Delta) = T_0 \Delta^{1-\gamma}$
\citep[e.g.,][]{hui97}.  However, as other authors have done, we will
assume a uniform UV background and an isothermal IGM
\citep{songaila02,songaila04,fan02,fan06}.  Following \citet{fan02},
we can then express the optical depth as a function of density,
\begin{equation}
   \tau(\Delta) = \tau_0 \left( \frac{1+z}{7} \right)^{4.5}
                         \left[ \frac{0.05}{\Gamma_{-12}(z)} \right]
                         \Delta^2 \, ,
   \label{eq:tau_delta}
\end{equation}
where $\Gamma_{-12}$ is the \hi\ ionization rate in units of
$10^{-12}~{\rm s}^{-1}$.  For comparison to other works
\citep{mcdonald01,fan02,fan06}, we take $\tau_0 = 82$, although the
normalization depends on the choice of cosmology.  Equations
(\ref{eq:mhr00_Pdelta}) and (\ref{eq:tau_delta}) can then be used to
determine the expected distribution of optical depths,
\begin{equation}
   P_{\tau}^{\rm MHR00}(\tau) = \frac{A G^{(\beta-1)/2}}{2 \tau^{(\beta+1)/2}}
                              \exp{\left[
                             -\frac{\left( G^{1/3}\tau^{-1/3}-C_0 \right)^2}
                               {2(2\delta_0/3)^2} \right]} \, ,
   \label{eq:mhr00_Ptau}
\end{equation}
where
\begin{equation}
   G \equiv \tau_0 \left(\frac{1+z}{7}\right)^{4.5} 
                   \left[\frac{0.05}{\Gamma_{-12}(z)}\right] \, .
   \label{eq:G_def}
\end{equation}
Finally, we can convert to the expected distribution of normalized
fluxes, $F = e^{-\tau}$,
\begin{equation}
   P_{F}^{\rm MHR00}(F) = \frac{A G^{(\beta-1)/2}}
                             {2 (-\ln{F})^{(\beta+1)/2} F}
                        \exp{ \left[ 
                     -\frac{\left( G^{1/3}(-\ln{F})^{-1/3}-C_0 \right)^2}
                           {2(2\delta_0/3)} \right] } 
   \label{eq:mhr00_PF}
\end{equation}
for $0 \le F \le 1$, $0$ otherwise.  The distribution of fluxes at a
particular $z$ is then fully specified by the ionization rate \G.

\subsubsection{Lognormal $\tau$ distribution}\label{sec:lognormal_distribution}

For the lognormal optical depth distribution, we make no assumptions
about the underlying density field, temperature, or ionization rate.
As discussed above, a lognormal distribution can be motivated either
from arguments about the evolution of an initially Gaussian density
field \citep{coles91,bi92} or by the central limit theorem.  Here we
consider it to be a generic model that may plausibly describe the
distribution of optical depths.  The lognormal distribution is
described by two parameters, $\mu = \langle \ln{\tau} \rangle$, and
$\sigma$, which is the standard deviation of $\ln{\tau}$,
\begin{equation}
   P_{\tau}^{\rm Lognormal}(\tau) = \frac{1}{\tau \sigma \sqrt{2\pi}}
                           \exp{ \left[ 
                            -\frac{(\ln{\tau}-\mu)^2}{2\sigma^2} \right] } \, .
   \label{eq:lognormal_Ptau}
\end{equation}
This gives an expected distribution of transmitted fluxes,
\begin{equation}
   P_{F}^{\rm Lognormal}(F) = \frac{1}{(-\ln{F}) F \sigma \sqrt{2\pi}}
                           \exp{ \left[
                            -\frac{(\ln{(-\ln{F})}-\mu)^2}{2\sigma^2} 
                            \right] }
   \label{eq:lognormal_PF}
\end{equation}
for $0 \le F \le 1$, $0$ otherwise.  There are obvious similarities
between the MHR00 and lognormal distributions, which should not be
surprising if they are both expected to at least roughly describe the
data.  We will examine the differences between the two cases more
closely in \S\ref{sec:lognormal_params}.

\subsection{Fitting the observed PDFs}\label{sec:fitting_pdfs}

In order to match the observed flux PDF, we must account for various
imperfections in the data.  The most important of these is noise in
the flux measurements, which will smooth out the PDF and create pixels
with $F < 0$ and $F > 1$.  We incorporate this effect by convolving
the ideal flux PDFs given by equations (\ref{eq:mhr00_PF}) and
(\ref{eq:lognormal_PF}) with a smoothing kernel constructed separately
for each flux bin.  (Numerically, the smoothing is performed on bins
much narrower than those used for the final PDFs).  The kernel for a
particular bin is a weighted sum of Gaussian kernels whose widths and
weights are determined from the distribution of formal flux errors of
pixels in that bin.  The result is typically a kernel with a narrow
core to account for pixels with low noise, and an extended tail for
noisier pixels.  This allows us to fit regions of the \lya\ forest
where the data quality is highly inhomogeneous.

Errors in the continuum level and the flux zero point will also affect
the the observed PDF.  A change in the continuum will cause the
observed PDF to be stretched or compressed in proportion to the flux
level.  An error in the zero point, which may result either from
imperfect sky subtraction or from spurious counts (i.e., cosmic rays)
improperly handled by the spectrum extraction or combination routines,
will also stretch or compress the observed PDF from the low-flux end.
In fitting the PDFs we consider two cases: first, where we assume
there are no errors in either the continuum or the zero point, and
second, where the continuum level and zero point are treated as free
parameters.  We define the preferred continuum and zero point levels
to be those which, if applied to the data, would allow the theoretical
distributions to produce the best.  However, when performing the fits,
the adjustments are applied to the models and not to the data.  The
continuum and zero points are treated independently, such that a
change in the zero point does not require a change in the continuum,
and visa versa.  We do not allow zero point corrections at $z < 3$,
where few pixels have zero flux.  This was found to have no
significant impact on the other parameters.

\begin{figure}
  \epsscale{1.0}
  \centering
  \plotone{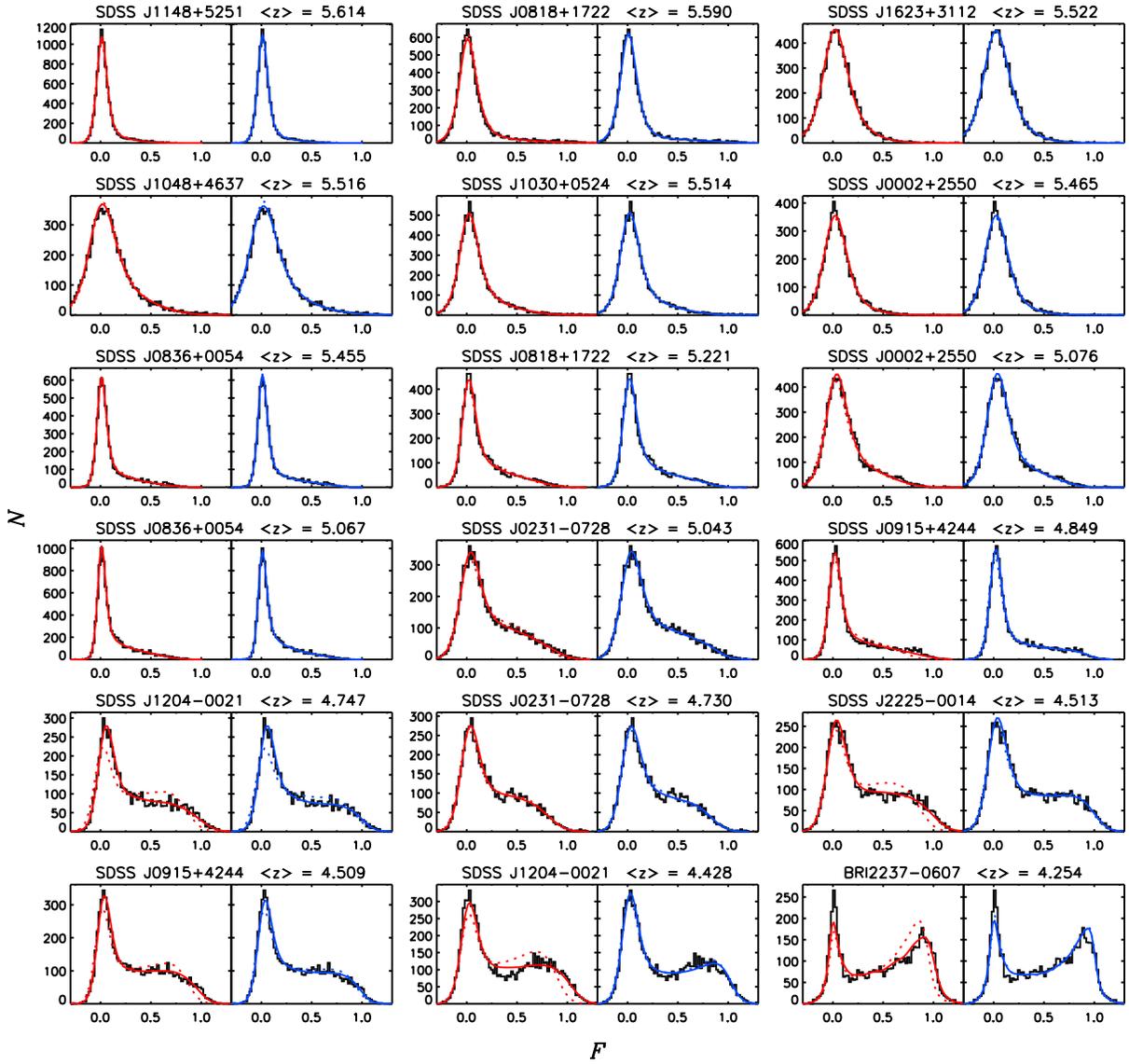}
  \caption[Fits to the \lya\ flux PDFs at $4.254 \le \langle z \rangle
  5.614$]{Fits to the \lya\ flux probability distribution functions for
    QSOs in our sample.  Each set of panels is labeled with the QSO
    name and the mean absorption redshift.  Histograms show the
    observed PDF.  For each region, MHR00 model fits assuming an
    isothermal IGM are shown on the left-hand side (red lines), while
    fits based on a lognormal $\tau$ distribution are shown on the
    right-hand side (blue lines).  Dotted lines indicate the best fit
    without adjusting either the continuum or the zero point.  Solid
    lines show the best fits when the continuum and zero point are
    allowed to vary.  The lognormal $\tau$ distribution generally
    produces a good fit without large adjustments to the continuum,
    whereas at $3 < z < 5$ the MHR00 model fits tend to be poor unless
    a significant continuum adjustment is made.\label{fig:pdf_fits_a}}
\end{figure}

\begin{figure}
  \epsscale{1.0}
  \centering
  \plotone{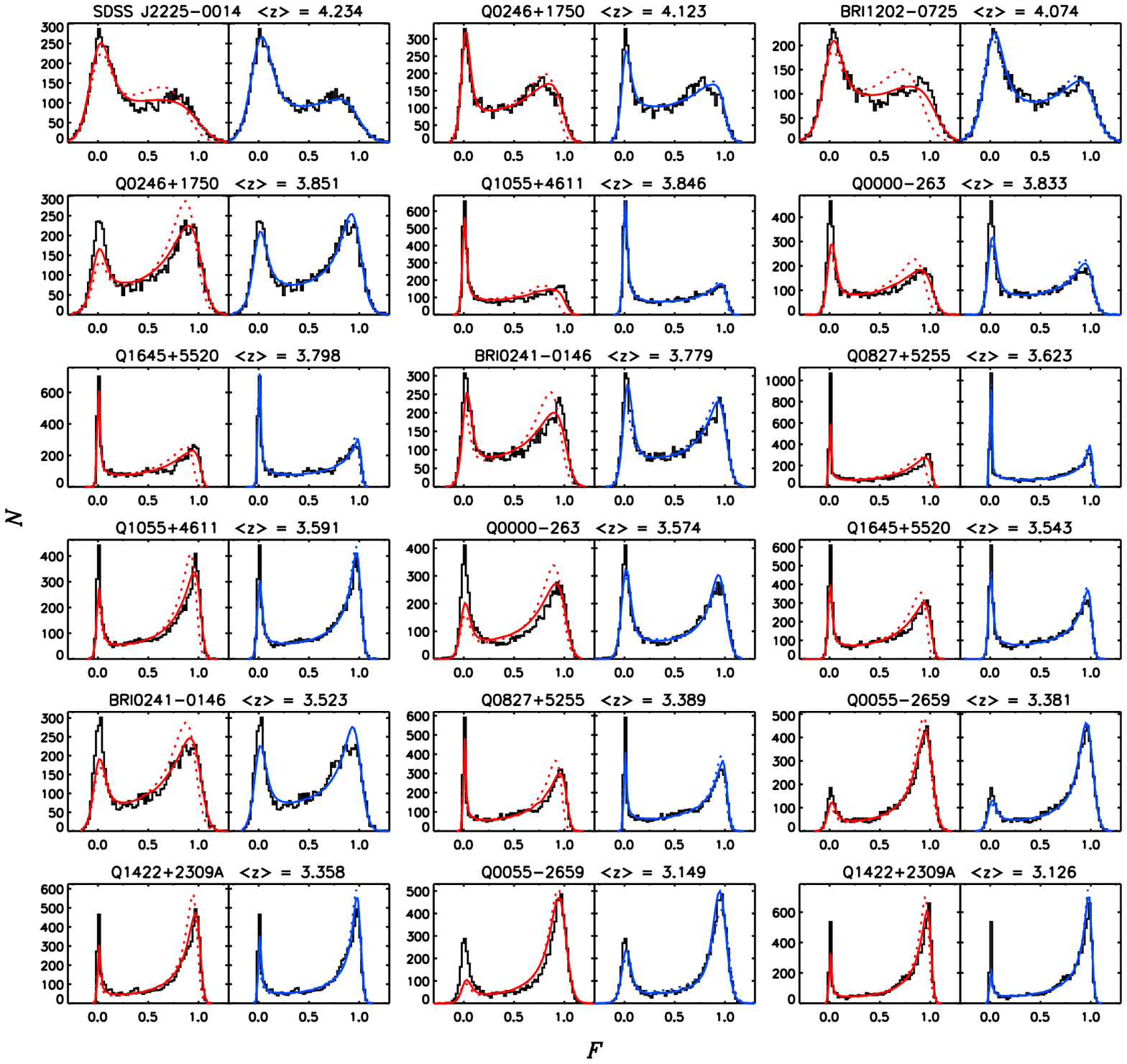}
  \caption[Fits to the \lya\ flux PDFs at $3.126 \le \langle z \rangle
  4.234$]{Fits to the \lya\ flux probability distribution functions for
    QSOs in our sample, continued from
    Figure~\ref{fig:pdf_fits_a}.\label{fig:pdf_fits_b}}
\end{figure}

\begin{figure}
  \epsscale{1.0}
  \centering
  \plotone{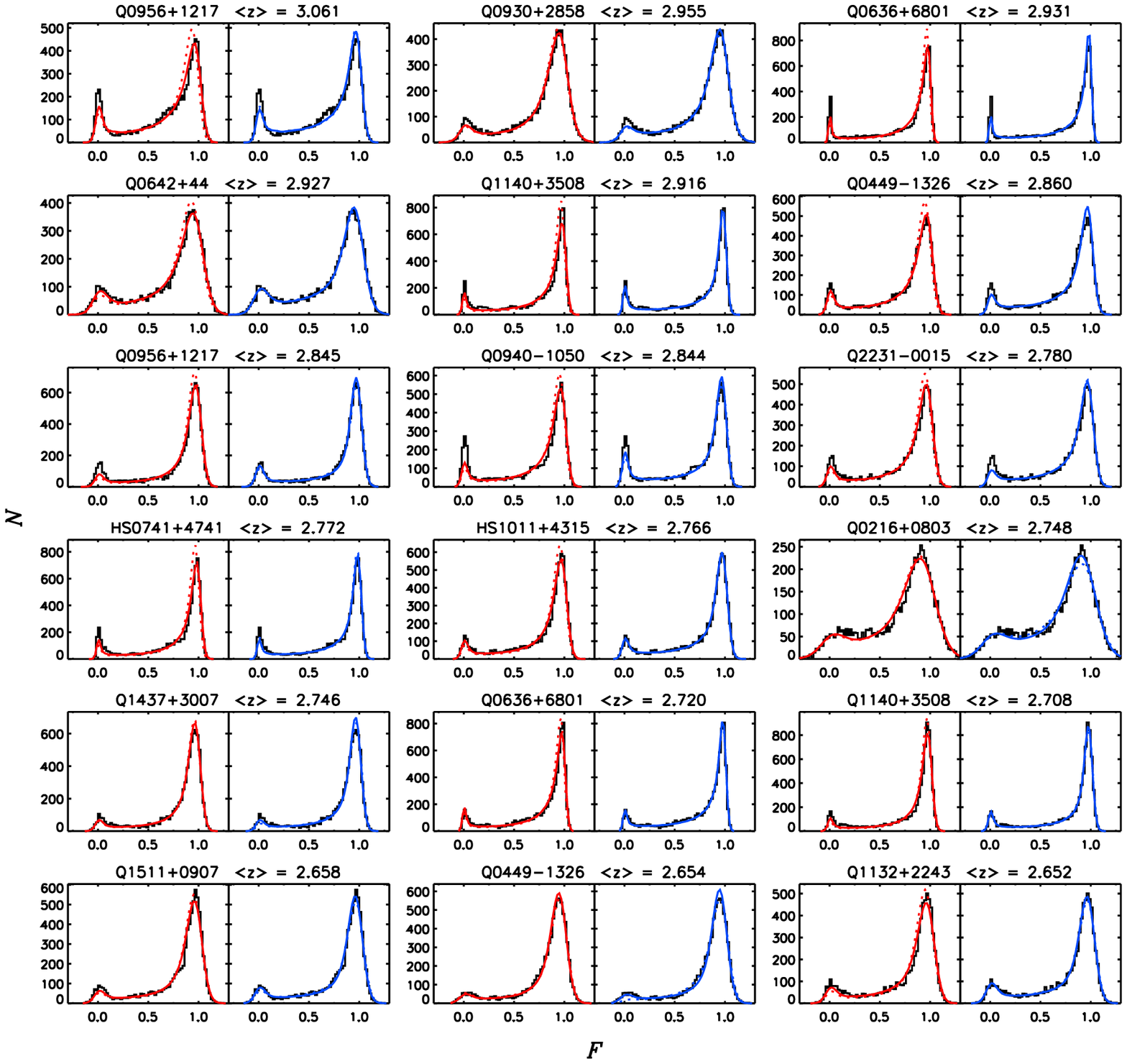}
  \caption[Fits to the \lya\ flux PDFs at $2.652 \le \langle z \rangle
  3.061$]{Fits to the \lya\ flux probability distribution functions for
    QSOs in our sample, continued from
    Figure~\ref{fig:pdf_fits_a}.\label{fig:pdf_fits_c}}
\end{figure}

\begin{figure}
  \epsscale{1.0}
  \centering
  \plotone{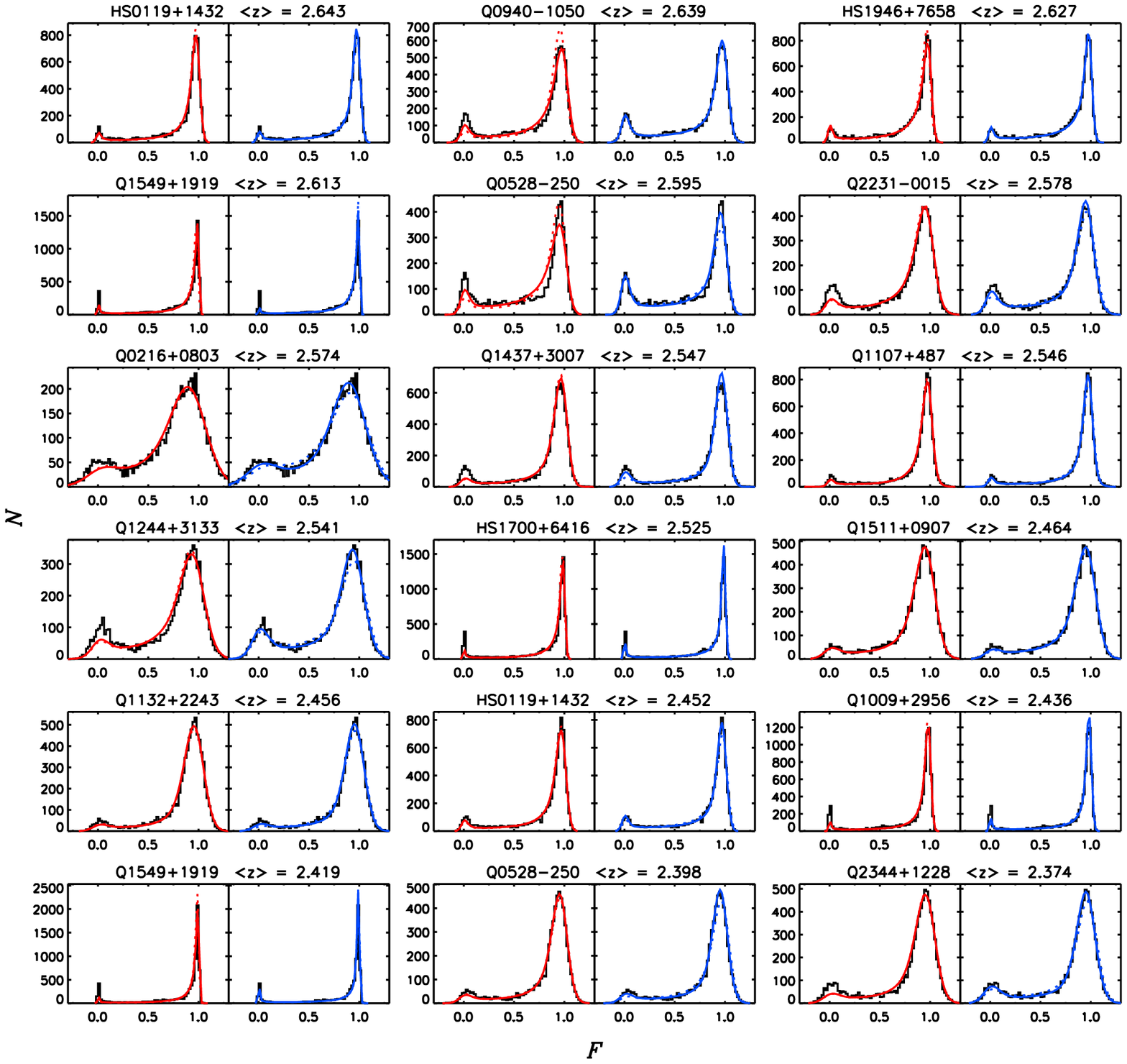}
  \caption[Fits to the \lya\ flux PDFs at $2.374 \le \langle z \rangle
  2.643$]{Fits to the \lya\ flux probability distribution functions for
    QSOs in our sample, continued from
    Figure~\ref{fig:pdf_fits_a}.\label{fig:pdf_fits_d}}
\end{figure}

\begin{figure}
  \epsscale{1.0}
  \centering
  \plotone{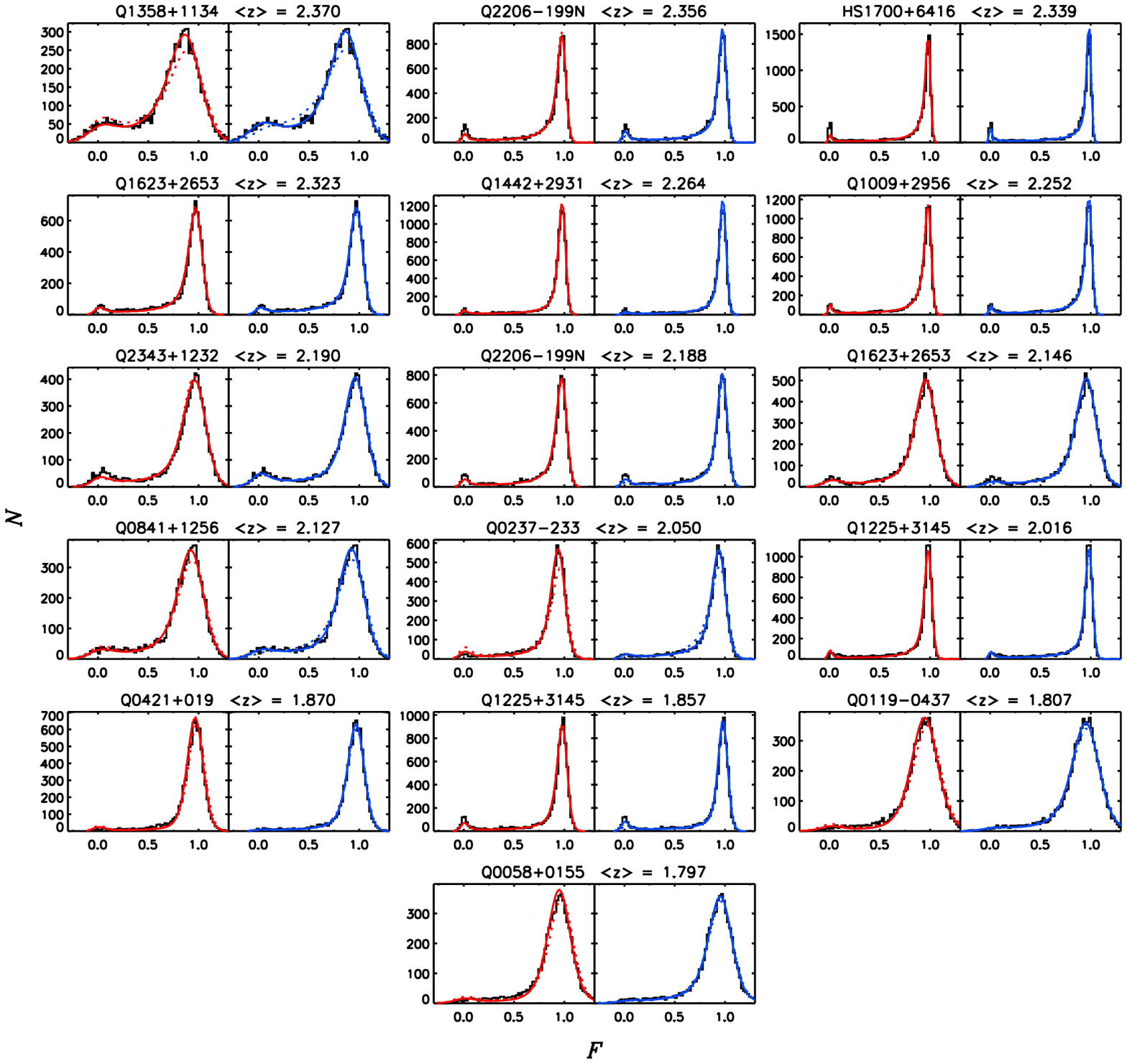}
  \caption[Fits to the \lya\ flux PDFs at $1.797 \le \langle z \rangle
  2.370$]{Fits to the \lya\ flux probability distribution functions for
    QSOs in our sample, continued from
    Figure~\ref{fig:pdf_fits_a}.\label{fig:pdf_fits_e}}
\end{figure}

The results of the \chisq\ minimization fitting are summarized in
Tables~\ref{tab:mhr00_params_isothermal} and
\ref{tab:lognormal_params} for the MHR00 and lognormal cases,
respectively.  The best-fitting PDFs are plotted in
Figures~\ref{fig:pdf_fits_a} through \ref{fig:pdf_fits_e}.  For each
region, we show the observed PDF along with the best-fitting
theoretical PDFs in the cases where no continuum or zero point
corrections are made and where the continuum and zero point are
allowed to vary.  At $z > 5$, the MHR00 and lognormal distribution
provide very similar fits.  This is not a surprise since, at these
redshifts, we are sampling the low-optical depth tail of both
distributions.  The differences in the distributions increase at lower
redshift.  At $3 < z < 5$, the best-fit MHR00 distribution
significantly under-predicts the number of pixels with very low optical
depth unless a continuum correction is applied.  In contrast, the
best-fit lognormal distributions provide a reasonable fit to the data
at all redshifts, with or without a change in the continuum.  Both
models under-predict the number of saturated pixels in some cases,
although the discrepancy tends to be much larger for the MHR00
distribution.

\begin{figure}
  \epsscale{1.0}
  \centering
  \plotone{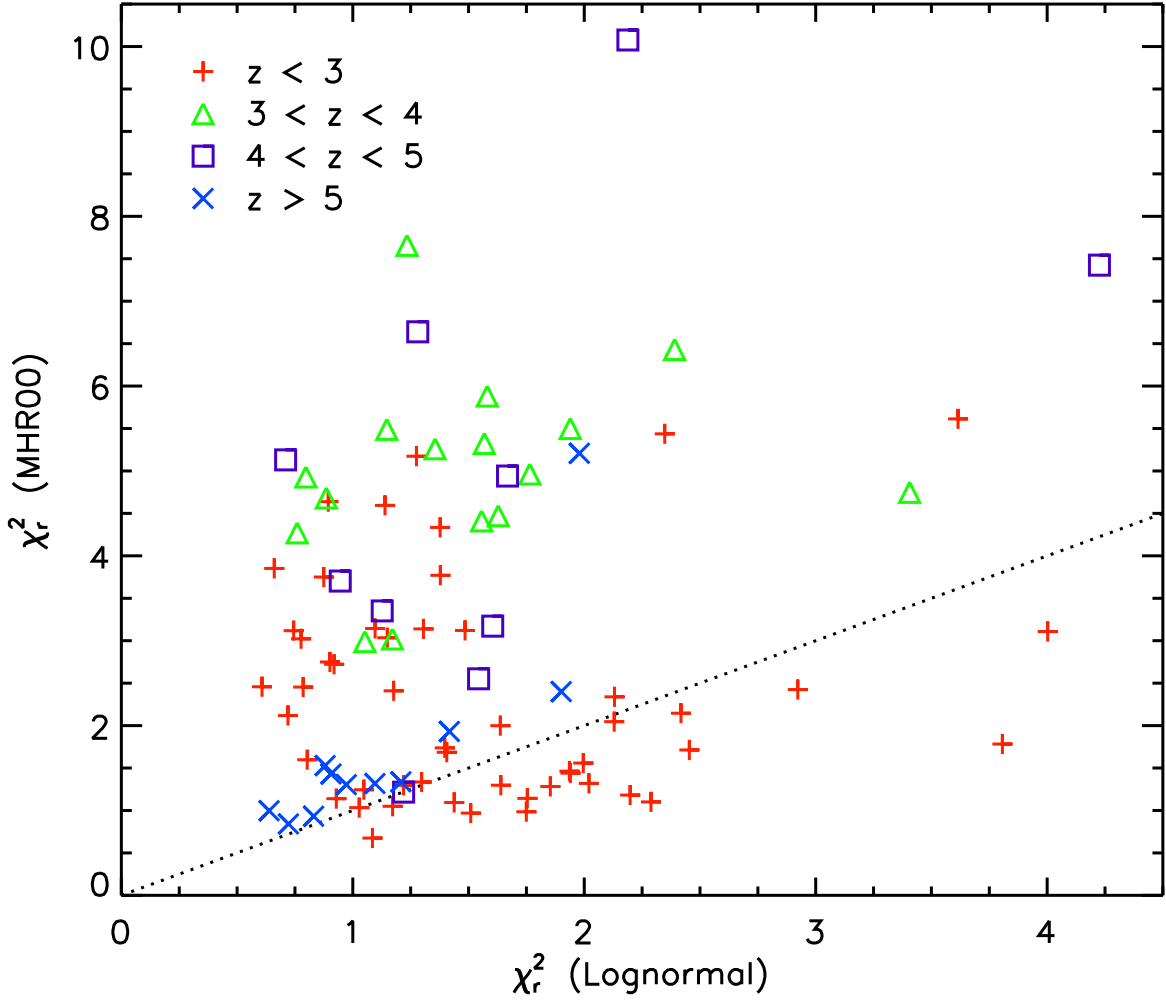}
  \caption[Reduced \chisq\ values for PDF fits: continuum and zero
  point fixed]{A comparison of the reduced \chisq\ values for the
    best-fitting MHR00 and lognormal $\tau$ PDFs when the continuum
    and zero point are held fixed.  Symbols indicate the mean
    absorption redshift of the fitted region of the \lya\ forest.  The
    dotted line indicates where $\chi^2_{\rm r,\,MHR00} = \chi^2_{\rm
      r,\,Lognormal}$.  Roughly half of the \lya\ regions at $z < 3$
    are better fit by the MHR00 PDF.  Otherwise, the lognormal PDF is
    preferred.\label{fig:chi_sq_fixed}}
\end{figure}

In Figure~\ref{fig:chi_sq_fixed} we compare the minimum reduced
\chisq\ values for both models in the case where the continua and zero
points are held fixed.  At $z < 2$, there is a roughly even divide
between regions that are better fit by the MHR00 distribution and
those that prefer the lognormal distribution.  However, in most
instances where the MHR00 distribution is preferred, the fit is
relatively poor ($\chi^2_{\rm r} > 2$).  At $z > 5$, the fits are
mostly comparable, as noted above.  For $3 < z < 4$, the lognormal
distribution provides a reasonable fit and is strongly preferred over
the MHR00 model.

\begin{figure}
  \epsscale{1.0}
  \centering
  \plotone{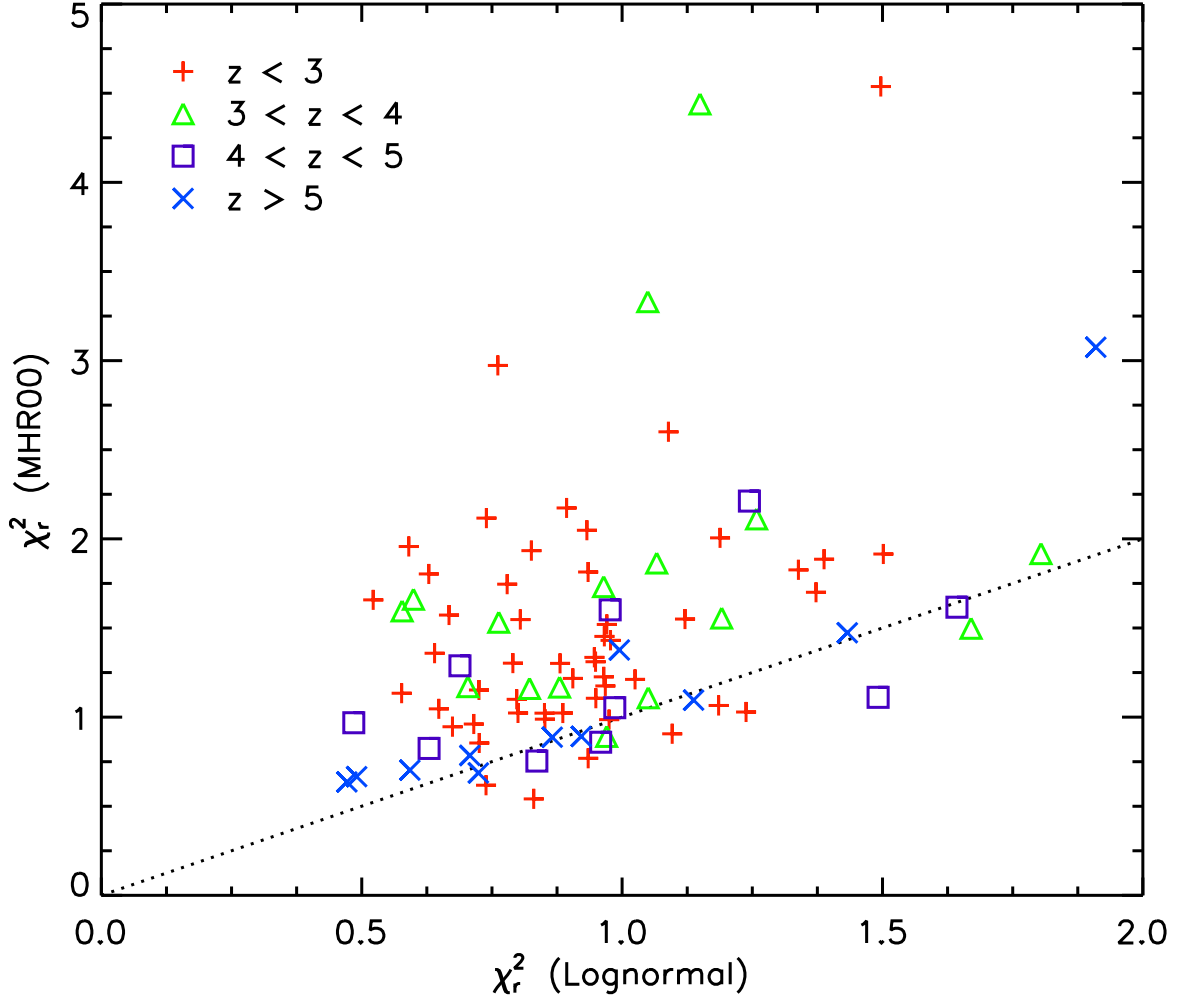}
  \caption[Reduced \chisq\ values for PDF fits: continuum and zero
  point allowed to vary]{A comparison of the reduced \chisq\ values
    for the best-fitting MHR00 and lognormal $\tau$ PDFs when the
    continuum and zero point are allowed to vary.  Note the change in
    scale from Figure~\ref{fig:chi_sq_fixed}.  Symbols indicate the
    mean absorption redshift of the fitted region of the \lya\ forest.
    The dotted line indicates where $\chi^2_{\rm r,\,MHR00} =
    \chi^2_{\rm r,\,Lognormal}$.  The lognormal PDF is preferred at
    all redshifts, particularly at $z < 5$.\label{fig:chi_sq_vary}}
\end{figure}

The fits improve for both models when the continua and zero points are
allowed to vary.  Most of this improvement is the result of the
continuum corrections.  The effect is particularly large for the MHR00
distribution, which implies that the MHR00 model tends to require that
a significant continuum correction be applied to the data in order to
produce a good fit.  In Figure~\ref{fig:chi_sq_vary} we plot the
reduced \chisq\ values for these more general fits.  As was the case
without continuum and offset adjustments, the two distributions
produce comparable fits at $z > 5$.  However, at all lower redshifts,
the lognormal distribution is preferred.

\begin{figure}
  \epsscale{1.0}
  \centering
  \plotone{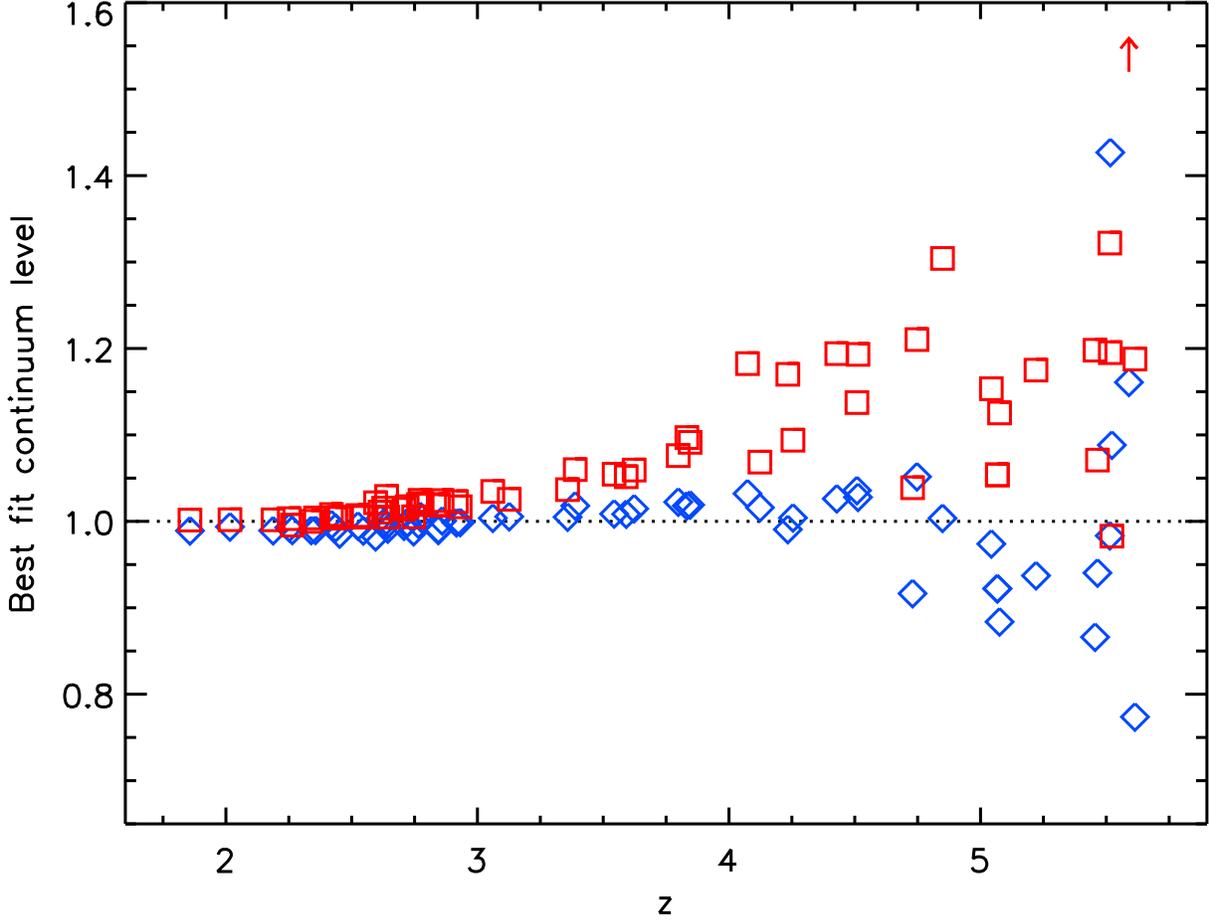}
  \caption[Continuum adjustments for PDF fits as a function of
  redshift]{Continuum adjustments to the data that are required so
    that the observed distribution of \lya\ fluxes are best fit by the
    theoretical distributions.  Squares show the continuum adjustment
    needed for the MHR00 model.  Diamonds show the continuum
    adjustment needed for the lognormal $\tau$ distribution.  The
    MHR00 model value for SDSS~J0818$+$1722 at $\langle z_{\rm abs}
    \rangle = 5.590$ lies outside the plot range, as indicated by the
    arrow.  At $z < 4$, only the values for regions with median flux
    error $< 0.05$ are shown.  The MHR00 distribution requires a
    steadily increasing continuum adjustment with redshift to account
    for the lack of pixels predicted to lie near the continuum.  At $z
    > 5.4$ the best-fitting continuum has a large scatter for both
    distributions due to how little transmitted flux remains in the
    forest.\label{fig:cont_vs_z}}
\end{figure}

As noted above, even at low redshift, where extended regions of the
spectrum have very little absorption, the continuum fit may be in
error due to a combination of noise and the personal bias of the
individual applying the fit.  However, at $z < 4$, the continuum error
should be less than a few percent for reasonably high signal-to-noise
data.  In Figure~\ref{fig:cont_vs_z} we plot the continuum correction
preferred for both distributions as a function of redshift.  The MHR00
model requires the continuum to steadily increase with redshift over
the continuum drawn by hand in order to account for the lack of pixels
predicted to lie near the continuum (i.e., pixels with very low
optical depth).  In contrast, the lognormal $\tau$ distribution
naturally accommodates fluxes near the continuum and does not require
a large continuum correction for $z < 4.5$.  At $z > 5.4$, the
preferred continuum adjustment has a large scatter for both
theoretical distributions, since nearly all pixels have significant
optical depth.

\begin{figure}
  \epsscale{1.0}
  \centering
  \plotone{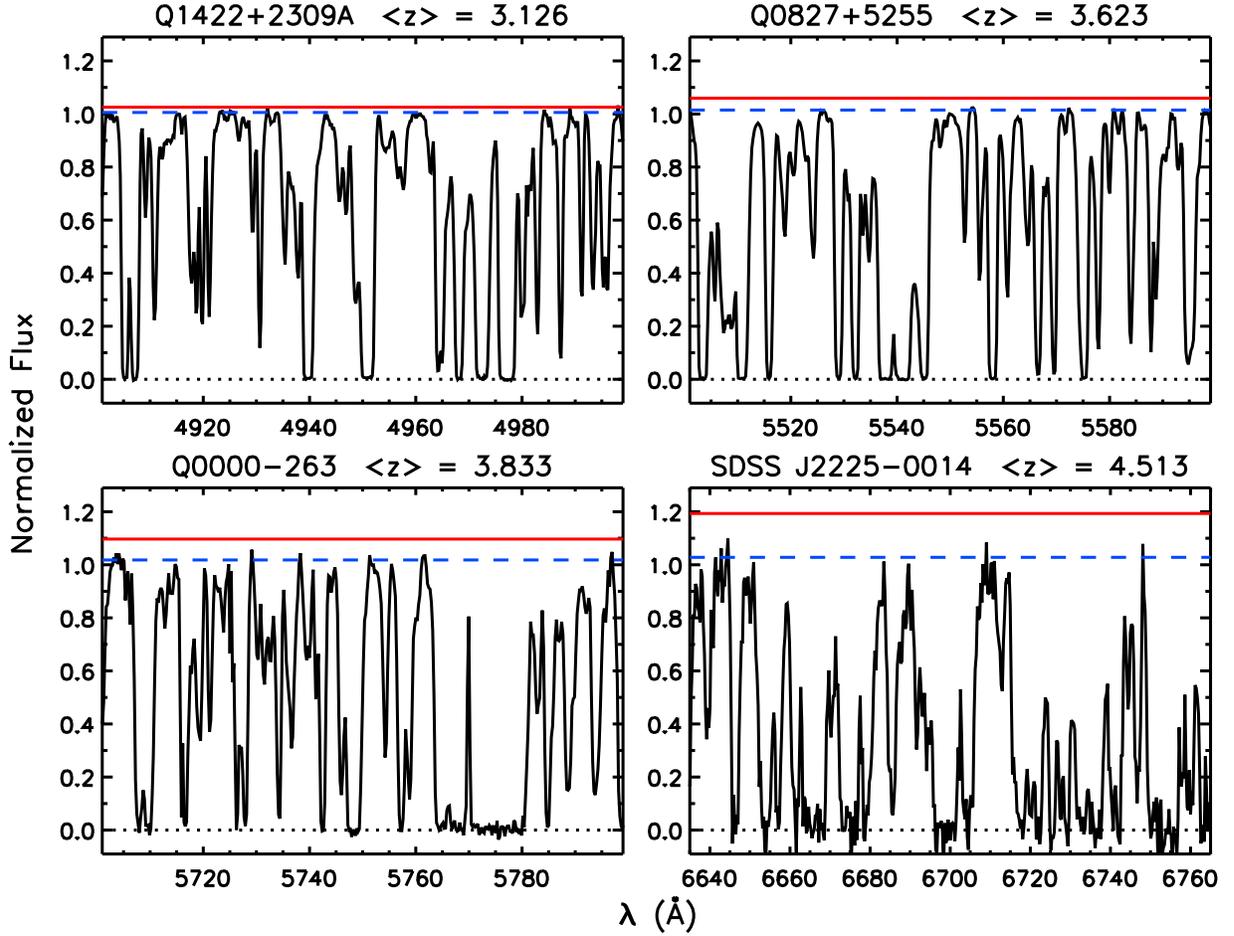}
  \caption[Examples of continuum adjustments overlaid on spectra]{Four
    examples of continuum adjustments needed so that the observed flux
    PDFs are best fit by the theoretical distributions.  Each panel
    shows a sample of the \lya\ forest taken from the fitted region
    indicated by the QSO name and mean absorption redshift.  The solid
    and dashed horizontal lines shows the continuum levels best fit by
    the MHR00 and lognormal $\tau$ distributions, respectively.  The
    spectra have been binned to 13~\kms\ pixels for
    clarity.\label{fig:cont_examples}}
\end{figure}

In Figure~\ref{fig:cont_examples} we show examples of the best-fit
continua overlaid on the corresponding regions of the \lya\ forest.
While the shape of a QSO continuum can be somewhat ambiguous when
convolved with the response function of the instrument, no undue
effort has been made to fit the continua across every transmission
peak.  The lognormal distribution fits the data well when the continua
are near their intuitive values, while the MHR00 model requires the
continua to be substantially higher.  Fitting QSO continua is an
inherently uncertain task.  However, even when the continuum is
allowed to vary, the lognormal $\tau$ distribution produces a better
fit than the MHR00 model.

In contrast to our results, \citet{rauch97} and \citet{mcdonald00}
found good agreement between the observed flux PDFs from some of the
same sightlines used here and the predictions from a numerical
simulation with a density distribution similar to the MHR00 model.
The reason for this appears to lie in their treatment of the
continuum.  Both works apply a strong correction to their simulated
spectra by placing the continuum at the maximum transmitted flux level
for each pass through the simulation box (10$h^{-1}$~Mpc, or $\sim
30$~\AA\ at $z = 4$).  This is a much higher-order correction than we
consider here.  In addition, \citet{mcdonald00} group all pixels with
flux $F > 1$ into their bin at $F = 1$.  This disguises the shape of
the observed PDF for pixels with low optical depth, particularly at $z
= 3-4$.  By fitting pixels at all fluxes, we remain sensitive to the
\textit{shape} of the PDF near $F \sim 1$.  Applying a low-order
continuum correction is therefore not sufficient to obtain a good fit
for the MHR00 distribution.  However, this works well in the lognormal
case.  Much of the discriminating power in the flux PDF occurs at very
low optical depths.  Therefore, unless more reliable continuum fits
can be made, the success of the MHR00 model in this regime is at best
unclear.

\section{Redshift Evolution of Optical Depth}\label{sec:redshift_evolution}

\subsection{Lognormal Parameters}\label{sec:lognormal_params}

\begin{figure}
  \epsscale{1.0}
  \centering
  \plotone{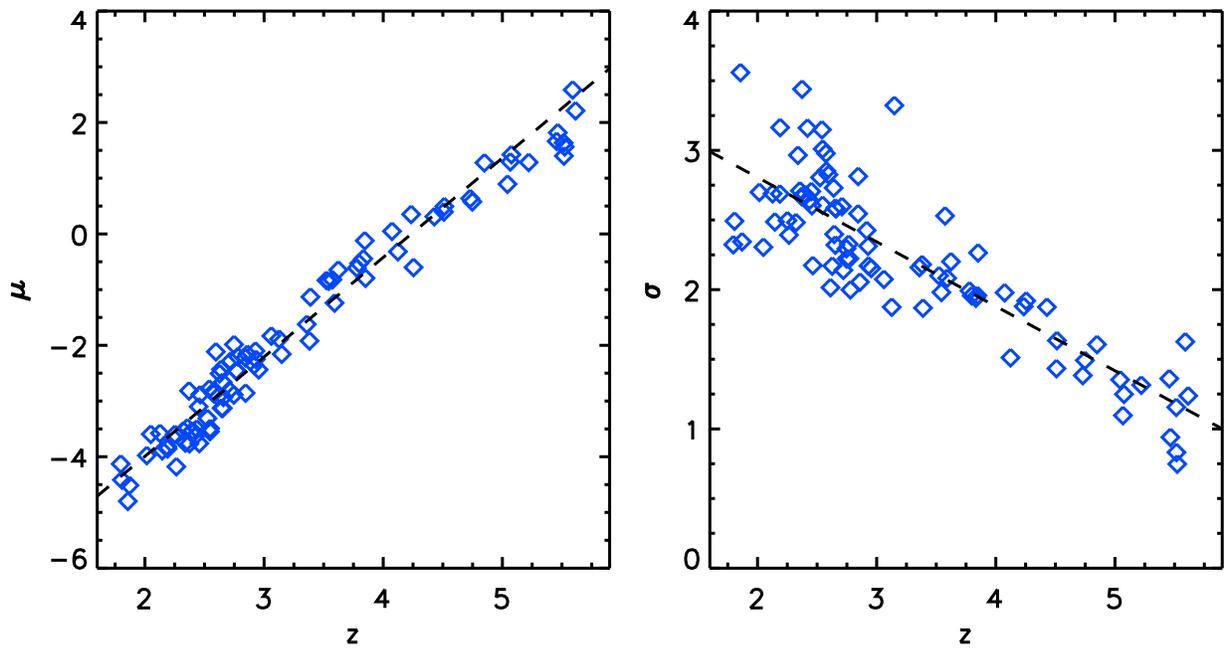}
  \caption[Lognormal $\tau$ distribution parameters]{Parameters for
    the lognormal distribution of \lya\ optical depths that produce
    the best fits to the observed flux PDFs.  Here, $\mu = \langle
    \ln{\tau} \rangle$ and $\sigma = $ std\,dev\,$(\ln{\tau})$.  The
    dashed lines show the best linear fits from equations
    (\ref{eq:mu_z}) and (\ref{eq:sig_z}).\label{fig:mu_sig_vs_z}}
\end{figure}

We have shown that a lognormal distribution of optical depths provides
a good fit to the observed \lya\ transmitted flux PDF at all redshifts
$1.7 < z < 5.8$.  In this section we examine the evolution of the
lognormal distribution and use it to predict the evolution of the mean
transmitted flux at $z > 6$.  In Figure~\ref{fig:mu_sig_vs_z} we plot
the lognormal parameters $\mu$ and $\sigma$ as a function of $z$.
Both parameters evolve smoothly with redshift, as should be expected
if they reflect a slowly-evolving density field, UV background, and
temperature-density relation.  The increase in $\mu$ and decrease in
$\sigma$ with $z$ can both be understood primarily in terms of the
evolution of a self-gravitating density field.  At earlier times, the
density contrast in the IGM will be lower.  This will tend to produce
a higher volume-weighted median $\tau$, which is given by $e^{\mu}$,
as well as a smaller logarithmic dispersion in $\tau$, which is given
by $\sigma$.  Since we do not have an {\it a priori} model for how the
lognormal parameters should evolve, for this work we choose the
simplest possible parametrization.  Excluding points at $z > 5.4$,
where the lognormal parameters depend on highly uncertain continuum
levels, a linear fit in redshift gives
\begin{equation}
   \mu(z) = (-9.35 \pm 0.17) + (1.79 \pm 0.04)(1+z) \, ,
   \label{eq:mu_z}
\end{equation}
\begin{equation}
   \sigma(z) = (4.19 \pm 0.16) - (0.46 \pm 0.03)(1+z) \, .
   \label{eq:sig_z}
\end{equation}
These fits are plotted as dashed lines in
Figure~\ref{fig:mu_sig_vs_z}.

\begin{figure}
  \epsscale{1.0}
  \centering
  \plotone{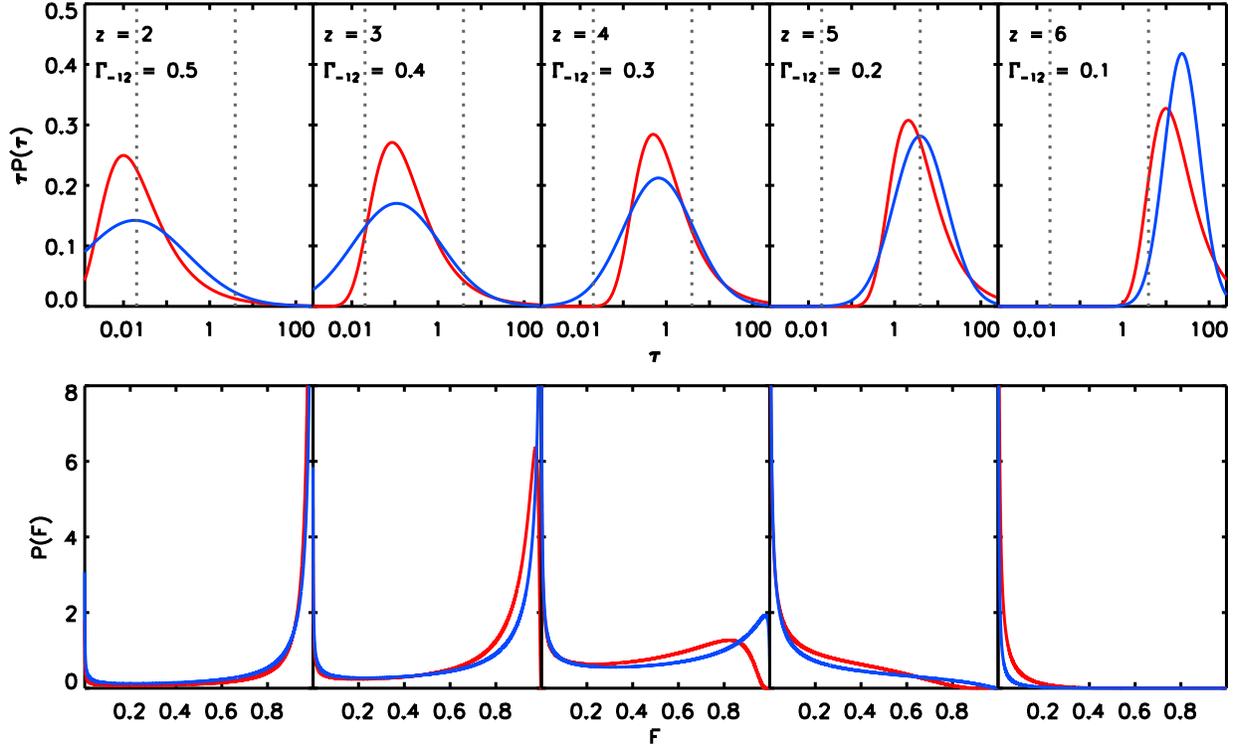}
  \caption[Redshift evolution of theoretical $\tau$ and flux
  distributions]{Redshift evolution of the theoretical \lya\ optical
    depth and transmitted flux distributions.  The top panels show the
    $\tau$ distributions for the indicated redshifts and ionization
    rates.  Bottom panels show the corresponding transmitted flux
    PDFs.  Distributions for the MHR00 model are shown in red.
    Distributions for the lognormal $\tau$ model are shown in blue.
    Parameters for the lognormal $\tau$ distribution were calculated
    from fits to $\mu$ and $\sigma$ as a function of redshift
    (cf. equations \ref{eq:mu_z} and \ref{eq:sig_z}).  Vertical dotted
    lines indicate optical depths corresponding to 98\% and 2\%
    transmitted flux.  The clearest differences in the predicted {\it
      shapes} of the flux PDFs occur at $3 < z < 5$.  The lognormal
    $\tau$ distribution, which produces better fits to the data,
    narrows with redshift more rapidly than the MHR00 $\tau$
    distribution.  Hence, fewer pixels with measurable transmitted
    flux at $z = 6$ are predicted in the lognormal
    case.\label{fig:tau_distributions}}
\end{figure}

We can compare the evolution of the MHR00 and lognormal $\tau$
distributions and their predictions for the transmitted flux PDF.  In
Figure~\ref{fig:tau_distributions} we plot fiducial $\tau$ and flux
distributions for $2 \le z \le 6$.  Parameters for the lognormal
distribution are calculated from equations (\ref{eq:mu_z}) and
(\ref{eq:sig_z}).  For the MHR00 model, values for \G\ are chosen to
be consistent with the fitted values in
Table~\ref{tab:mhr00_params_isothermal}.  The vertical dotted lines
indicate the range of optical depths that can be measured with good
data.  At $z = 2$ we are primarily sensitive to the high-$\tau$ tail
in both distributions.  At higher redshifts, the peaks of the
distributions shift towards higher values of $\tau$ until we are
sampling only the end of the low-$\tau$ tail at $z = 6$.

Differences in the shape of the transmitted flux PDF are largest at $3
\le z \le 5$, where $P_{\tau}(\tau)$ is well-sampled.  The fact that
the lognormal $\tau$ distribution is most strongly favored at these
redshifts suggests that it is more likely to be useful in making
predictions for the distribution of transmitted flux at $z > 6$.  An
important feature of the lognormal distribution is that it narrows
with redshift more rapidly than the MHR00 distribution.  It therefore
predicts fewer pixels with measurable transmitted flux at $z \sim 6$
than does the MHR00 model with a slowly evolving UV background.

\subsection{Mean transmitted flux}\label{sec:mean_transmitted_flux}

\begin{figure}
  \epsscale{1.0}
  \centering
  \plotone{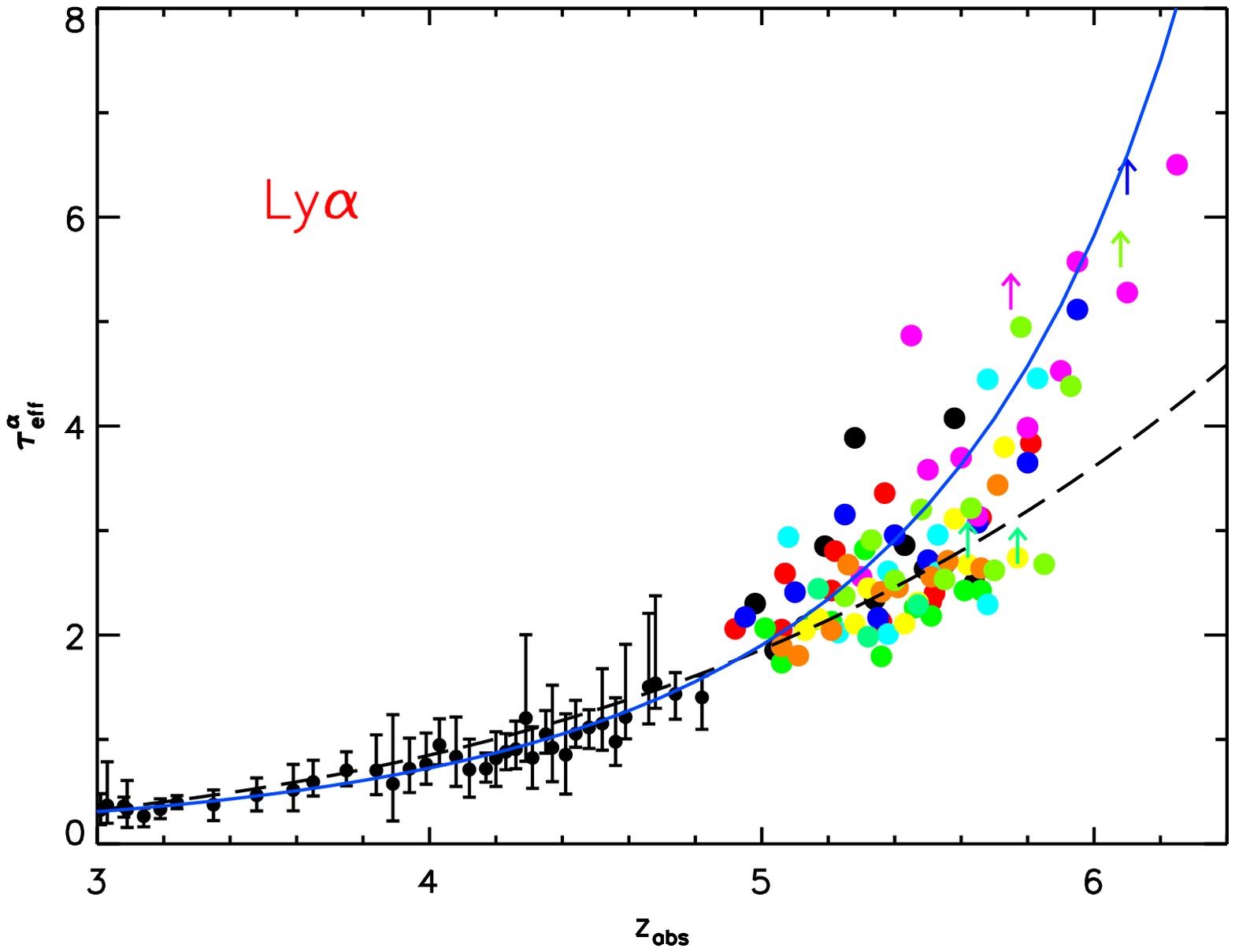}
  \caption[Evolution of \taueffa\ over $3 \le z \le 6.2$ (linear)]{The
    evolution of \lya\ effective optical depth with redshift, where
    $\tau_{\rm eff} = -\ln{\langle F \rangle}$.  Data points are from
    \citet{songaila04} (small circles) and \cite{fan06} (large circles
    and arrows, with colors matching their Figure~2).  The dashed line
    shows the best-fit power-law to \taueffa\ at $z < 5.5$ from
    \citet{fan06}.  The solid line shows \taueffa\ calculated from the
    lognormal distribution of \lya\ optical depths, for which the
    parameters were fit at $z < 5.4$.  A simple evolution in the
    lognormal $\tau$ distribution predicts the upturn in \taueffa\ at
    $z > 5.5$ and produces a better fit to the observed \taueffa\ at
    $4 < z < 5$.\label{fig:taueff_lya}}
\end{figure}

\begin{figure}
  \epsscale{1.0}
  \centering
  \plotone{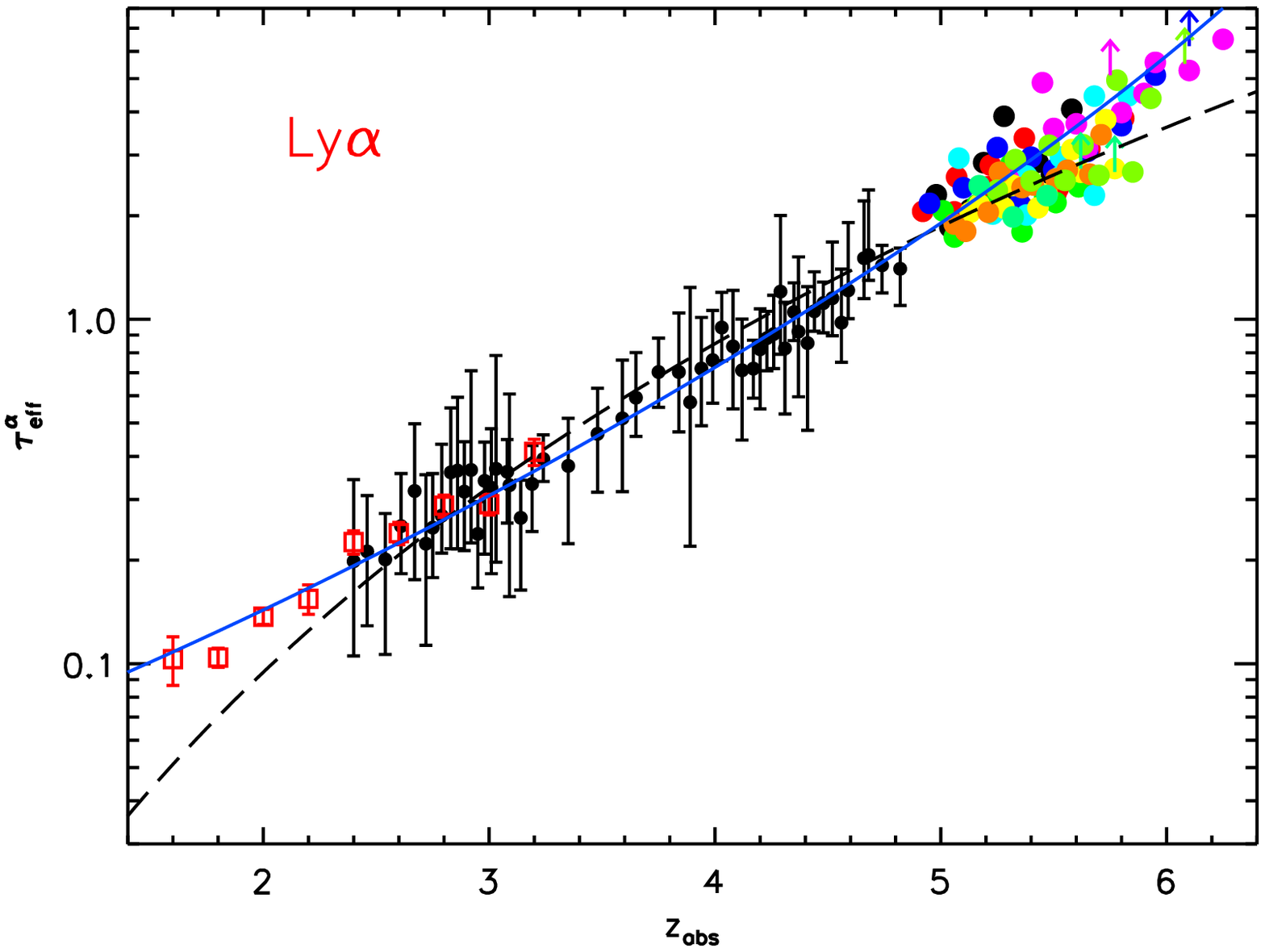}
  \caption[Evolution of \taueffa\ over $1.6 \le z \le 6.2$
  (logarithmic)]{Same as Figure~\ref{fig:taueff_lya}, with $\tau_{\rm
      eff}^{\alpha}$ on a logarithmic scale.  We have also included
    lower-redshift measurements calculated from \citet{kirkman05},
    which exclude absorption from metal lines, Lyman limit systems,
    and damped \lya\ systems.  The \citet{kirkman05} points are
    plotted as open squares with errors in the mean measurements.  The
    power-law fit from \citet{fan06} (dashed line) under-predicts the
    amount of \lya\ absorption both at $z > 5.7$ and at $z < 2.5$.  In
    contrast, $\tau_{\rm eff}^{\alpha}$ calculated from the lognormal
    $\tau$ distribution (solid line), provides a simultaneously good
    fit to all points at $1.6 < z < 6.2$.\label{fig:taueff_lya_log}}
\end{figure}

We can use the redshift evolution of the lognormal distribution to
predict the the evolution of transmitted flux at $z \gtrsim 6$.  The
mean transmitted flux will be given by
\begin{equation}
  \langle F \rangle = \int_0^1 F \, P_F(F)\, dF \, .
  \label{eq:mean_F}
\end{equation}
It is conventional to express the mean flux in terms of an effective
optical depth $\tau_{\rm eff} = -\ln{\langle F \rangle}$.  For a
distribution of optical depths, $\tau_{\rm eff}$ will be smaller than
the true mean optical depth.  We show measurements of $\tau_{\rm eff}$
for \lya\ from \citet{songaila04} and \citet{fan06} in
Figure~\ref{fig:taueff_lya}.  The dashed line shows the best-fitting
power-law to their data at $z < 5.5$ from \citet{fan06}.  The
deviation of the data from the power-law at $z > 5.7$ has been cited
as the primary evidence for an abrupt change in the ionizing
background at $z \sim 6$.  We also show $\tau_{\rm eff}^{\alpha}$ as
predicted by the evolution of the lognormal $\tau$ distribution given
by equations (\ref{eq:mu_z}) and (\ref{eq:sig_z}) as a solid line.  We
emphasize that the lognormal parameters were fit only to measurements
at $z < 5.4$.  Even so, $\tau_{\rm eff}^{\alpha}$ calculated from the
lognormal distribution both better fits the data at $z < 5$ and {\it
  predicts} the upturn in $\tau_{\rm eff}^{\alpha}$ at $z > 5.7$.  In
Figure~\ref{fig:taueff_lya_log} we include the lower-redshift
measurements of \citet{kirkman05}.  The power-law under-predicts the
amount of \lya\ absorption at $z < 2.5$, while the lognormal
distribution matches all observations at $1.6 < z < 6.2$.

\begin{figure}
  \epsscale{1.0}
  \centering
  \plotone{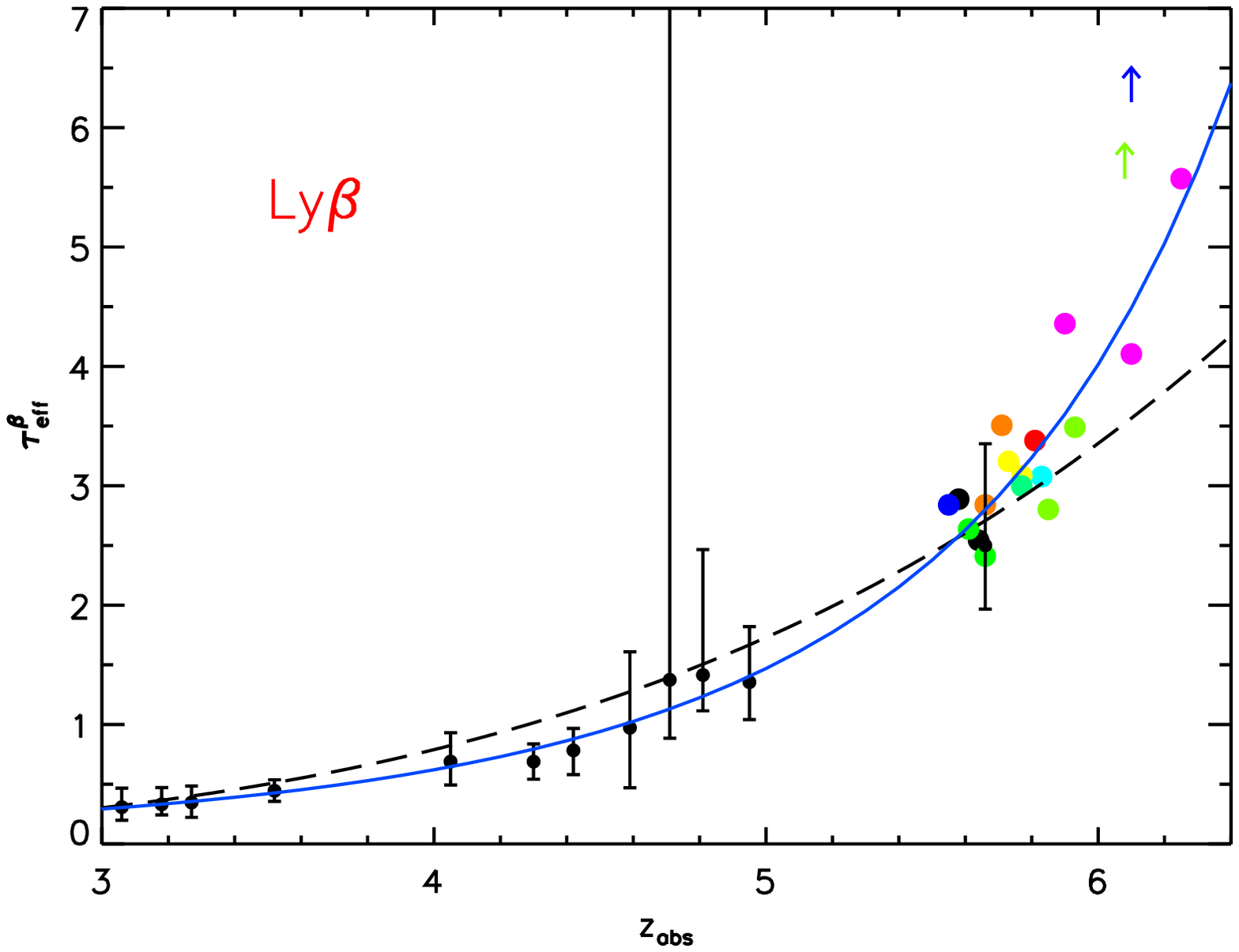}
  \caption[Evolution of \taueffb\ over $3 \le z \le 6.2$ (linear)]{The
    evolution of \lyb\ effective optical depths with redshift.  Data
    points at from \citet{songaila04} (small circles) and \cite{fan06}
    (large circles and arrows, with colors matching their Figure~3).
    The data have {\it not} been corrected for foreground \lya\
    absorption.  The dashed line shows the best-fit power-law to
    \taueffb\ at $z < 5.5$ from \citet{fan06}.  The solid line shows
    \taueffb\ predicted purely from the lognormal distribution of
    \lya\ optical depths.  Even though no independent fitting of \lyb\
    fluxes was performed, the lognormal $\tau$ distribution captures
    the upturn in \taueffb\ at $z > 5.5$ and produces a better fit to
    the observed \taueffb\ at $4 < z < 5$.\label{fig:taueff_lyb}}
\end{figure}

Stronger constraints on the ionization state of the IGM can be set
using \lyb, which is a weaker transition than \lya\ by a factor of
6.2.  In the lognormal case, this produces a distribution of \lyb\
optical depths with the same $\sigma$ as \lya\ but with $\mu_{\beta} =
\mu_{\alpha} - \ln{6.2}$.  We can then compute the expected mean flux
in the \lyb\ forest at redshift $z$ by multiplying the mean
transmission resulting from \lyb\ absorption at $z$ by the mean
transmission resulting from \lya\ absorption at $z_{\beta} =
(1+z)\lambda_{\beta}/\lambda_{\alpha} - 1$.  We show the $\tau_{\rm
  eff}^{\beta}$ measurements from \citet{songaila04} and \citet{fan06}
in Figure~\ref{fig:taueff_lyb}.  These are computed directly from the
transmitted flux and have not been corrected for foreground \lya\
absorption.  The dashed line again shows the best-fit power-law to the
points at $z < 5.5$ from \citet{fan06}.  The solid line shows the
lognormal prediction.  Here again, despite the fact that we have not
used any \lyb\ measurements to determine the optical depth
distribution, $\tau_{\rm eff}^{\beta}$ predicted in the lognormal case
is a better fit to the data at $z < 5$ and follows the upturn in
$\tau_{\rm eff}^{\beta}$ at $z > 5.7$.

Our purpose here is not to fully characterize the evolution of
transmitted flux at all redshifts.  We have simply identified a
distribution of optical depths that describes the observed
distribution of transmitted fluxes better than the commonly used
model.  The fact that this distribution evolves smoothly with
redshift, and that the same evolution describes changes in the \lya\
forest as well at $z \sim 6$ as it does at $z \sim 3$ strongly
suggests that the disappearance of transmitted flux at $z > 6$ is due
to a smooth evolution of IGM properties.  The lognormal prediction for
$\tau_{\rm eff}^{\beta}$ falls slightly below some of the lower limits
of \citet{fan06} at $z \sim 6$, but the prediction does not take into
account the expected scatter in the mean flux or any small deviation
from our adopted linear redshift evolution of the lognormal
parameters.  The important point is that the evolution of the mean
transmitted flux can be well described by a smooth evolution in the
underlying optical depths.  When sampling only the tail of the $\tau$
distribution, as at $z \sim 6$, a slight change in the optical depths
will produce a large change in the transmitted flux.

\subsection{UV background}\label{uv_background}

\citet{liu06} recently demonstrated that a semi-analytic model based
on a lognormal {\it density} distribution can reproduce the observed
rise in $\tau_{\rm eff}$ at $z > 5.7$.  However, they invoke a UV
background that declines rapidly with redshift, decreasing by a factor
of $\sim 11$ from $z = 3$ to $5$, and by a factor of $\sim 7$ from $z
= 5$ to $6$.  We have not assumed that the lognormal $\tau$
distribution used here arises directly from a lognormal density
distribution.  However, if we assume a uniform UV background and an
isothermal IGM, than we can calculate the \hi\ ionization rate by
inverting equation (\ref{eq:tau_delta}) and averaging over all
densities.  Doing so gives
\begin{equation}
   \Gamma_{-12} = 0.05 \left( \frac{1+z}{7} \right)^{4.5}
                      \frac{\tau_0}{\langle \tau^{1/2} \rangle ^2} \, ,
   \label{eq:fluxpdf_Gamma}
\end{equation}
where $\langle \tau^{1/2} \rangle^2 = e^{\mu + \sigma^2 / 4}$, and we
have used the fact that $\langle \Delta \rangle = 1$.  

\begin{figure}
  \epsscale{1.0}
  \centering
  \plotone{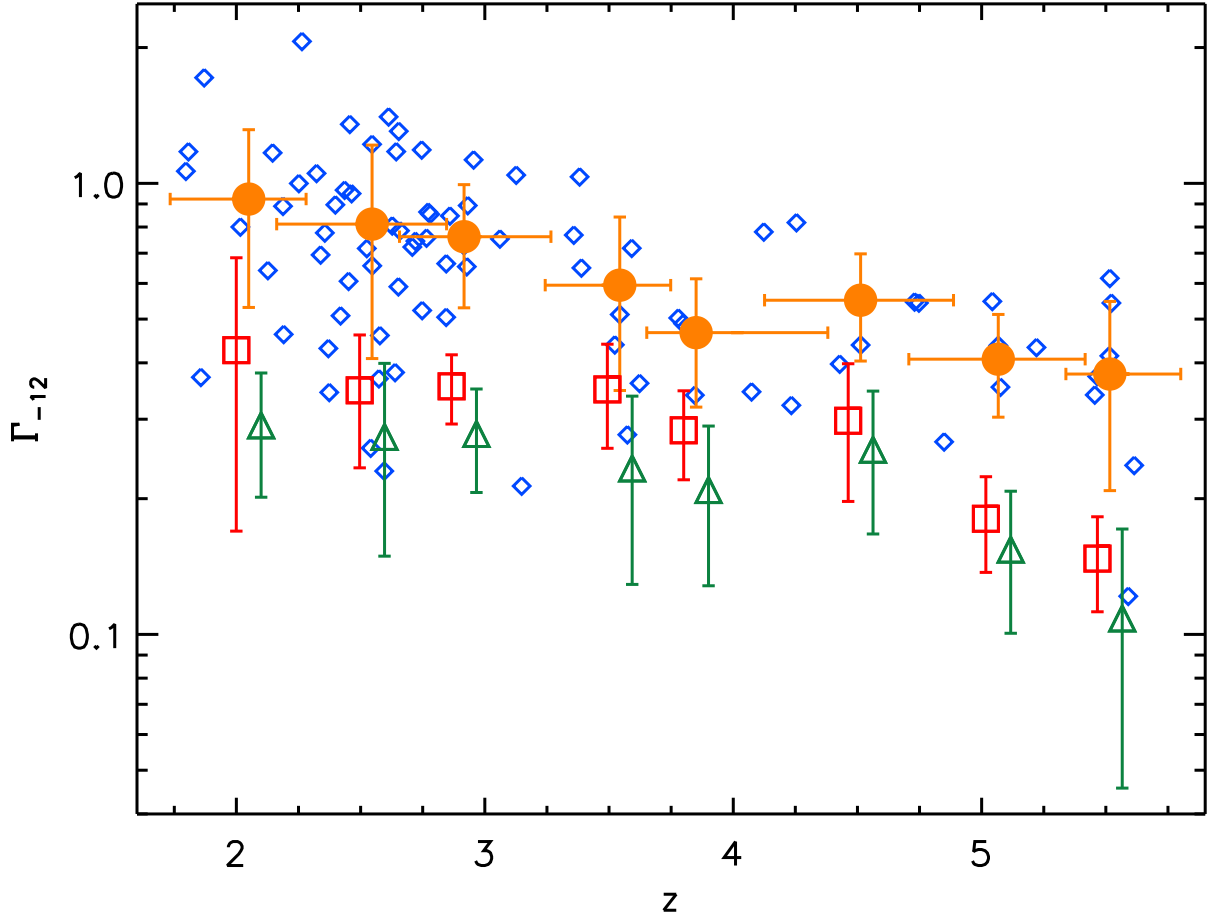}
  \caption[\hi\ ionization rates determined from flux PDFs]{\hi\
    ionization rate as a function of redshift.  Small diamonds show
    \G\ calculated from the lognormal fits to individual regions,
    assuming a uniform UV background (cf. equation
    \ref{eq:fluxpdf_Gamma}).  Points fit to high-$S/N$ data are shown
    in blue.  Filled circles show the mean \G\ for the lognormal model
    in redshift bins of 0.5, starting at $z = 2$.  Vertical error bars
    show the standard deviation of points within a bin.  Horizontal
    error bars show the range of redshift covered by all points within
    that bin.  Squares show the mean \G\ from the MHR00 model fits for
    an isothermal IGM and uniform UV background ($\alpha = 0$).
    Triangles show the mean \G\ from the MHR00 model fits when
    $\alpha$ is allowed to vary.\label{fig:gamma_vs_z}}
\end{figure}

In Figure~\ref{fig:gamma_vs_z} we show \G\ calculated for each fitted
region along with the mean values in bins of redshift.  For
comparison, the best-fit values of \G\ for the model distribution are
also shown.  The lognormal values are somewhat higher than the model
values, which are in turn roughly consistent with previous
measurements \citep{mcdonald01,fan06}.  However, we do not require the
strong evolution in \G\ given by \citet{liu06} for the lognormal
model.  Transforming from densities to optical depths depends on a
number of factors, and we do not presume that the assumptions implicit
in equation (\ref{eq:fluxpdf_Gamma}) are valid.  We merely point out
that a lognormal $\tau$ distribution is consistent with a slowly
evolving UV background.

We can also calculate the mean volume-weighted neutral fraction,
\begin{equation}
   f_{\mhi} = (5.5 \times 10^{-5}) \, h_{70}^{-1} 
                       \left( \frac{\Omega_{\rm m}}{0.3} \right)^{1/2}
                       \left( \frac{\Omega_{\rm b}}{0.04} \right)^{-1}
                       (1+z)^{-3/2} \langle \tau \rangle \, ,
   \label{eq:neutral_frac}
\end{equation}
where we have used $H(z) \approx H_0\Omega_{\rm m}^{1/2}(1+z)^{3/2}$.
The mean optical depth for the lognormal distribution will be $\langle
\tau \rangle = e^{\mu + \sigma^2 / 2}$.  Calculating $\mu$ and
$\sigma$ from equations (\ref{eq:mu_z}) and (\ref{eq:sig_z}), this
gives $f_{\mhi} = [1.0,1.2,1.9,4.0,11,20] \times 10^{-5}$ for $z =
[2,3,4,5,6,6.5]$.  The mean optical depth will depend strongly on the
high-$\tau$ tail of the distribution, which is poorly constrained at
$z > 4$.  However, the disappearance of transmitted flux at $z > 6$ is
at least consistent with a highly-ionized IGM.


\section{An inverse temperature-density relation?}\label{sec:temp_density}

We have shown that the simplest transformation of the MHR00 gas
density distribution to optical depths provides at best an uncertain
fit to the observed distribution of transmitted fluxes.  However,
there are several ways to modify the expected $\tau$ distribution.
Here we consider a non-isothermal temperature-density relation.  From
equation (\ref{eq:tau_delta_propto}) we have $\tau \propto
T^{-0.7}\Gamma^{-1}\Delta^2$.  We will address the general case where
either $T$ or $\Gamma$ may depend on $\Delta$.  For a power law
$T^{0.7}\Gamma \propto \Delta^{\alpha}$, this gives
\begin{equation}
   \tau(\Delta) = \tau_0 \left( \frac{1+z}{7} \right)^{4.5}
                         \left[ \frac{0.05}{\Gamma_{-12}(z)} \right]
                         \Delta^{2-\alpha} \, ,
   \label{tau_temp_density}
\end{equation}
where $\Gamma_{-12}$ is now the \hi\ ionization rate at the mean
density, and the temperature at the mean density is included in
$\tau_0$.  For a uniform UV background, the equation of state index
will be $\gamma = 1 + 1.43\alpha$.

\begin{figure}
  \epsscale{0.95}
  \centering
  \plotone{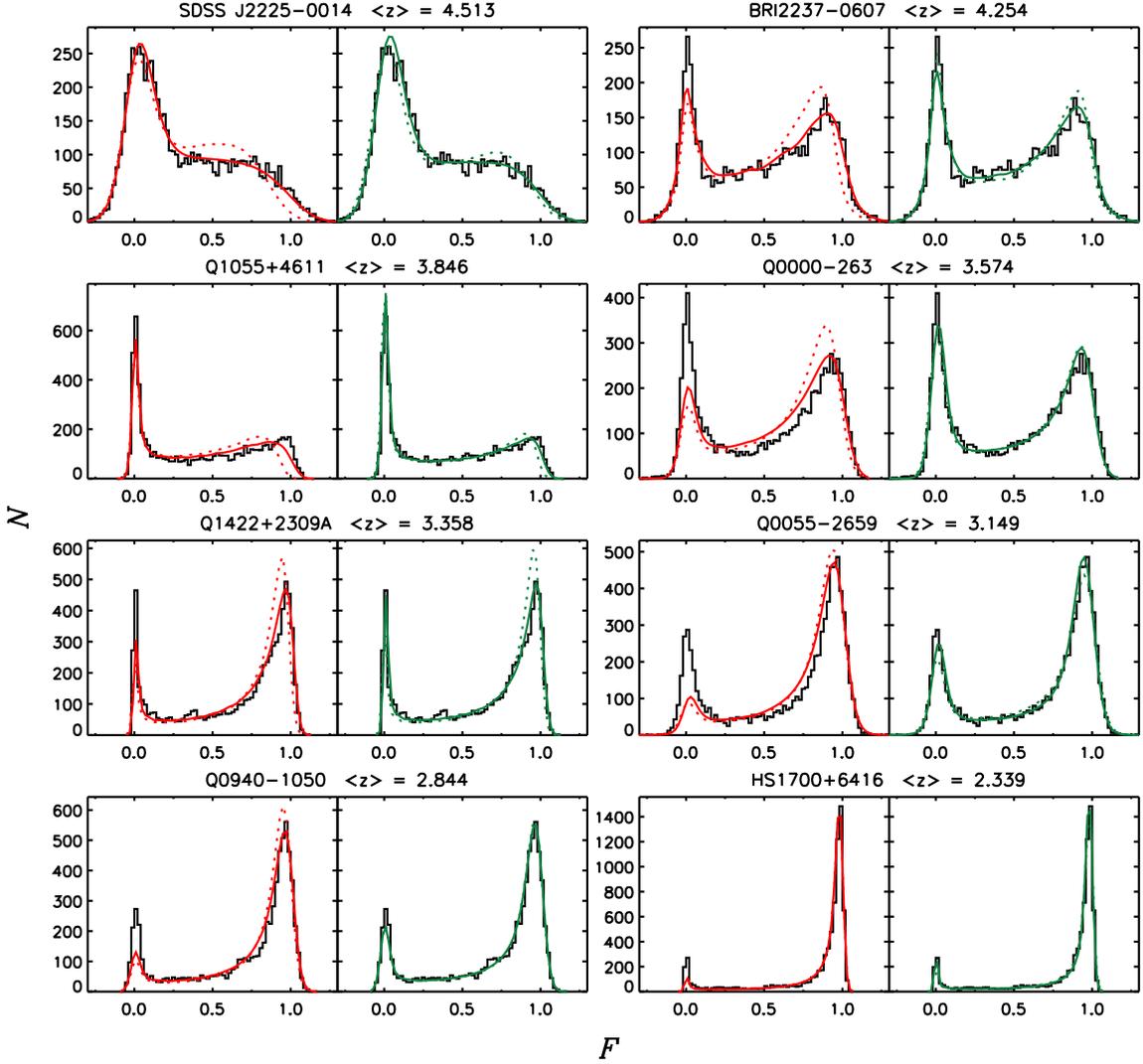}
  \caption[Example fits to \lya\ flux PDFs using a non-isothermal
  MHR00 model]{Examples of transmitted flux PDFs where the MHR00 model
    fit is significantly improved by allowing a non-isothermal
    temperature-density relation, or more generally, $T^{0.7} \Gamma
    \propto \Delta^{\alpha}$. Each set of panels is labeled with the
    QSO name and the mean absorption redshift.  Histograms show the
    observed PDF.  For each section, MHR00 model fits with $\alpha =
    0$ are shown on the left-hand side (red lines), while fits with
    $\alpha$ treated as a free parameter are shown on the right-hand
    side (green lines).  Dotted lines indicate the best fit without
    adjusting either the continuum or the zero point.  Solid lines
    show the best fits when the continuum and zero point are allowed
    to vary.  The mean value of $\alpha$ for all regions is $\langle
    \alpha \rangle \approx -0.4$.  This may indicate that the UV
    background decreases with density, or that there exists an inverse
    temperature-density relation.  Alternatively, finding $\alpha < 0$
    may be an artifact of some other features of the MHR00 model that
    causes it to disagree with the data.\label{fig:alpha_model_fits}}
\end{figure}

Not surprisingly, adding a degree of freedom significantly improves
the fits for many of our \lya\ forest regions.  The fitting results
are summarized in Table~\ref{tab:mhr00_params_nonisothermal}, and a
sample of the fits are shown in Figure~\ref{fig:alpha_model_fits}.
There is a large scatter in the best-fit $\alpha$ at all redshifts
when the continuum and zero point are allowed to vary.  However, the
mean value $\langle \alpha \rangle = -0.36 \pm 0.45$ (sample variance)
suggests that $T^{0.7}\Gamma$ {\it increases} towards lower densities.
For a uniform UV background, this implies an equation of state
$T(\Delta) \propto \Delta^{\gamma-1}$ with $\gamma \approx 0.5$.  An
index $< 1$ disagrees with previous measurements using the flux PDF
\citep{choudhury01,lidz06b,desjacques05}.  However, those works
typically considered only $\gamma > 1$, which is expected following
reionization if overdense regions experience more photoionization
heating and less adiabatic cooling than underdense regions.  Radiative
transfer effects may create a complex temperature-density relation if
underdense regions are reionized by a harder UV background than the
dense regions near ionizing sources \citep{bolton04}.  For the flux
PDF, $\gamma < 1$ allows for a lower $\Gamma$ (typically by $\sim
20$\%, see Figure~\ref{fig:gamma_vs_z}), creating more saturated
pixels, while at the same time maintaining a low $\tau$ in low density
regions.  The necessary continuum corrections also decrease, although
they are still roughly half of those needed in the case of $\alpha =
0$.  Of course, it is possible that we are not measuring the real
equation of state, and that the added degree of freedom simply
compensates for some other aspect of the model distribution.  A more
careful treatment of this problem will be reserved for future work.

\section{Conclusions}\label{sec:fluxpdf_conclusions}

We have analyzed the \lya\ transmitted flux probability distribution
in a high-resolution sample of 63 QSOs spanning the absorption
redshift range $1.7 < z < 5.8$.  Our main goal has been to assess how
well the theoretical optical depth distribution commonly used to
measure the \hi\ ionization rate describes the observed flux PDF.  We
find that the MHR00 model, under the assumptions of a uniform UV
background and an isothermal IGM, produces a poor fit to the observed
flux PDF at all redshifts where the optical depth distribution is well
sampled.  This discrepancy eases only if large continuum corrections
are applied.

In contrast, a lognormal distribution of optical depths fits the data
well with only minor continuum adjustments.  The parameters of the
lognormal distribution evolve smoothly with redshift, as expected for
a slowly evolving IGM, and reflect both an increase in the mean $\tau$
and a decrease in the relative scatter in $\tau$ with redshift.  We
have performed simple linear fits to the lognormal parameters at $z <
5.4$.  The mean transmitted flux calculated from these fits matches
the observations at $1.6 < z < 5.7$ better than the best-fitting power
law \citep{fan06}.  In addition, extrapolating the lognormal evolution
to $z > 6$ predicts the observed upturn in both \lya\ and \lyb\
effective optical depths.  This strongly suggests that if a slowly
evolving density field, ionizing background, and IGM temperature are
responsible for the evolution of the \lya\ forest at $z < 5$, then
there is no reason to suspect a sudden change in the IGM at $z \sim
6$.

We emphasize that we have used the lognormal distribution as a
phenomenological description of the optical depths only, and that the
distribution may not hold for optical depths that are outside the
dynamic range of the transmitted flux.  Other factors, such a
non-isothermal IGM or variations in the UV background are likely to be
important in deriving the optical depth distribution from the
underlying density field.  We have explored the possibility of a
non-isothermal IGM in the context of the MHR00 model.  The best fits
tend to favor an inverse temperature-density relation, where
temperature increases with density.  This is contrary to typical
expectations for the balance between photoionization heating and
adiabatic cooling \citep{hui97}, and may be an artifact of some other
feature that causes the MHR00 model to disagree with the data.
However, as \citet{bolton04} point out, radiative transfer effects may
create a complex thermodynamic state in the IGM.  If gas at a given
density can have a range of temperatures and/or ionization rates, then
a MHR00-like density distribution may give rise to a $\tau$
distribution that is closer to lognormal.

The largest source of uncertainty in fitting the flux PDFs remains the
continuum level.  Much of the disagreement between the MHR00 model and
observed PDFs stems from the lack of pixels predicted to have very low
optical depths at $z > 3$.  This can be at least partially remedied by
adjusting the continuum \citep[see also][]{mcdonald00}.  However, a
dramatic change in the IGM would still be required to explain the
observed lack of transmitted flux at $z \sim 6.2$
\citep[e.g.,][]{fan02,fan06}.  Future observation of $z > 4$ gamma-ray
bursts, whose continuum is a simple power law, may help to establish
the correct flux PDF.  For now, we have identified an optical depth
distribution that both fits the data down to $z = 1.6$ and captures
the evolution of the mean transmitted flux at $z > 5.7$.  If the
lognormal distribution truly reflects aspects of the real optical
depth distribution, then the motivation for late reionization may be
greatly diminished.

\acknowledgments

The authors would like to thank Martin Haehnelt for stimulating
conversations, as well as Tom Barlow and Rob Simcoe for reducing much
of the data.  We especially thank the Hawaiian people for the
opportunity to observe from from Mauna Kea.  Without their hospitality
this work would not have been possible.  MR has been supported by the
NSF under grant AST 05-06845.

\clearpage
   
\begin{deluxetable}{lccccc}
   \tabletypesize{\scriptsize}
   \tablewidth{0pt}
   \centering
   \tablecaption{Fitted \lya\ Forest Regions}
   \tablehead{
        \colhead{QSO} & \colhead{$z_{\rm QSO}$} & 
        \colhead{$\langle z_{\rm abs} \rangle$\tablenotemark{a}} & 
        \colhead{$z_{\rm abs}^{\rm min}$} &
        \colhead{$z_{\rm abs}^{\rm max}$} &
        \colhead{median flux error}
   }
   \startdata
   SDSS~J1148$+$5251 &  6.42 & 5.614 & 5.430 & 5.802 & 0.05 \\
   SDSS~J1030$+$0524 &  6.30 & 5.514 & 5.339 & 5.692 & 0.11 \\
   SDSS~J1623$+$3112 &  6.25 & 5.522 & 5.339 & 5.709 & 0.15 \\
   SDSS~J1048$+$4637 &  6.23 & 5.516 & 5.339 & 5.696 & 0.15 \\
   SDSS~J0818$+$1722 &  6.00 & 5.590 & 5.417 & 5.766 & 0.10 \\
                     &       & 5.221 & 5.066 & 5.417 & 0.07 \\
   SDSS~J0002$+$2550 &  5.82 & 5.465 & 5.339 & 5.592 & 0.12 \\
                     &       & 5.076 & 4.910 & 5.245 & 0.11 \\
   SDSS~J0836$+$0054 &  5.80 & 5.455 & 5.339 & 5.573 & 0.05 \\
                     &       & 5.067 & 4.893 & 5.245 & 0.04 \\
   SDSS~J0231$-$0728 &  5.42 & 5.043 & 4.885 & 5.206 & 0.11 \\
                     &       & 4.730 & 4.563 & 4.885 & 0.09 \\
   SDSS~J0915$+$4244 &  5.20 & 4.849 & 4.707 & 4.993 & 0.07 \\
                     &       & 4.509 & 4.373 & 4.647 & 0.07 \\
   SDSS~J1204$-$0021 &  5.09 & 4.747 & 4.582 & 4.887 & 0.09 \\
                     &       & 4.428 & 4.277 & 4.582 & 0.09 \\
   SDSS~J2225$-$0014 &  4.87 & 4.513 & 4.381 & 4.647 & 0.10 \\
                     &       & 4.234 & 4.087 & 4.381 & 0.11 \\
      BRI1202$-$0725 &  4.69 & 4.074 & 3.929 & 4.214 & 0.10 \\
      BRI2237$-$0607 &  4.56 & 4.254 & 4.126 & 4.377 & 0.07 \\
        Q0246$+$1750 &  4.44 & 4.123 & 3.988 & 4.260 & 0.05 \\
                     &       & 3.851 & 3.716 & 3.988 & 0.07 \\
        Q1055$+$4611 &  4.15 & 3.846 & 3.719 & 3.975 & 0.02 \\
                     &       & 3.591 & 3.460 & 3.719 & 0.03 \\
         Q0000$-$263 &  4.13 & 3.833 & 3.704 & 3.961 & 0.05 \\
                     &       & 3.574 & 3.447 & 3.704 & 0.05 \\
        Q1645$+$5520 &  4.10 & 3.798 & 3.672 & 3.927 & 0.01 \\
                     &       & 3.543 & 3.417 & 3.672 & 0.02 \\
      BRI0241$-$0146 &  4.08 & 3.779 & 3.652 & 3.906 & 0.05 \\
                     &       & 3.523 & 3.398 & 3.652 & 0.06 \\
        Q0827$+$5255 &  3.91 & 3.623 & 3.503 & 3.748 & 0.01 \\
                     &       & 3.389 & 3.265 & 3.503 & 0.02 \\
        Q0055$-$2659 &  3.65 & 3.381 & 3.266 & 3.499 & 0.05 \\
                     &       & 3.149 & 3.033 & 3.266 & 0.06 \\
       Q1422$+$2309A &  3.63 & 3.358 & 3.243 & 3.475 & 0.02 \\
                     &       & 3.126 & 3.011 & 3.243 & 0.02 \\
        Q0930$+$2858 &  3.44 & 2.955 & 2.845 & 3.067 & 0.07 \\
          Q0642$+$44 &  3.40 & 2.927 & 2.818 & 3.037 & 0.08 \\
        Q0956$+$1217 &  3.31 & 3.061 & 2.954 & 3.169 & 0.04 \\
                     &       & 2.845 & 2.738 & 2.954 & 0.05 \\
       HS0741$+$4741 &  3.23 & 2.772 & 2.664 & 2.876 & 0.03 \\
        Q0636$+$6801 &  3.18 & 2.931 & 2.827 & 3.036 & 0.02 \\
                     &       & 2.720 & 2.618 & 2.827 & 0.02 \\
        Q1140$+$3508 &  3.16 & 2.916 & 2.813 & 3.021 & 0.03 \\
                     &       & 2.708 & 2.605 & 2.813 & 0.03 \\
       HS1011$+$4315 &  3.14 & 2.766 & 2.657 & 2.869 & 0.04 \\
        Q0449$-$1326 &  3.10 & 2.860 & 2.757 & 2.962 & 0.04 \\
                     &       & 2.654 & 2.552 & 2.757 & 0.07 \\
        Q0940$-$1050 &  3.08 & 2.844 & 2.743 & 2.947 & 0.04 \\
                     &       & 2.639 & 2.538 & 2.743 & 0.05 \\
       HS1946$+$7658 &  3.07 & 2.627 & 2.524 & 2.728 & 0.03 \\
        Q2231$-$0015 &  3.02 & 2.780 & 2.680 & 2.881 & 0.05 \\
                     &       & 2.578 & 2.479 & 2.680 & 0.07 \\
         Q1107$+$487 &  2.98 & 2.546 & 2.446 & 2.646 & 0.04 \\
        Q1437$+$3007 &  2.98 & 2.547 & 2.448 & 2.648 & 0.05 \\
        Q0216$+$0803 &  2.98 & 2.574 & 2.487 & 2.658 & 0.15 \\
        Q1437$+$3007 &  2.98 & 2.746 & 2.648 & 2.846 & 0.04 \\
        Q0216$+$0803 &  2.98 & 2.748 & 2.659 & 2.843 & 0.12 \\
        Q1244$+$3133 &  2.97 & 2.541 & 2.439 & 2.638 & 0.10 \\
        Q1511$+$0907 &  2.89 & 2.658 & 2.562 & 2.756 & 0.06 \\
                     &       & 2.464 & 2.368 & 2.562 & 0.08 \\
        Q1132$+$2243 &  2.88 & 2.652 & 2.556 & 2.750 & 0.06 \\
                     &       & 2.456 & 2.361 & 2.556 & 0.09 \\
       HS0119$+$1432 &  2.87 & 2.643 & 2.547 & 2.740 & 0.03 \\
                     &       & 2.452 & 2.353 & 2.547 & 0.04 \\
        Q1549$+$1919 &  2.84 & 2.613 & 2.517 & 2.707 & 0.01 \\
                     &       & 2.419 & 2.324 & 2.517 & 0.01 \\
         Q0528$-$250 &  2.81 & 2.595 & 2.492 & 2.683 & 0.05 \\
                     &       & 2.398 & 2.302 & 2.492 & 0.07 \\
        Q2344$+$1228 &  2.79 & 2.374 & 2.280 & 2.470 & 0.09 \\
       HS1700$+$6416 &  2.74 & 2.525 & 2.432 & 2.619 & 0.01 \\
                     &       & 2.339 & 2.244 & 2.432 & 0.02 \\
        Q1442$+$2931 &  2.66 & 2.264 & 2.169 & 2.352 & 0.03 \\
        Q1009$+$2956 &  2.65 & 2.436 & 2.345 & 2.527 & 0.02 \\
                     &       & 2.252 & 2.162 & 2.343 & 0.02 \\
        Q1358$+$1134 &  2.58 & 2.370 & 2.282 & 2.461 & 0.15 \\
        Q2343$+$1232 &  2.58 & 2.190 & 2.101 & 2.281 & 0.10 \\
        Q2206$-$199N &  2.57 & 2.356 & 2.269 & 2.447 & 0.03 \\
                     &       & 2.188 & 2.105 & 2.269 & 0.04 \\
        Q1623$+$2653 &  2.53 & 2.323 & 2.235 & 2.411 & 0.06 \\
                     &       & 2.146 & 2.058 & 2.235 & 0.10 \\
        Q0841$+$1256 &  2.51 & 2.127 & 2.038 & 2.214 & 0.12 \\
         Q0237$-$233 &  2.24 & 2.050 & 1.966 & 2.128 & 0.06 \\
        Q1225$+$3145 &  2.21 & 2.016 & 1.938 & 2.098 & 0.03 \\
                     &       & 1.857 & 1.777 & 1.938 & 0.04 \\
         Q0421$+$019 &  2.05 & 1.870 & 1.795 & 1.947 & 0.08 \\
        Q0119$-$0437 &  1.98 & 1.807 & 1.733 & 1.876 & 0.14 \\
        Q0058$+$0155 &  1.96 & 1.797 & 1.734 & 1.859 & 0.12
   \enddata
   \tablenotetext{a}{Mean absorption redshift.}
   \label{tab:fitted_regions}
\end{deluxetable}

\clearpage

\begin{deluxetable}{lccccccccc}
   \tabletypesize{\scriptsize}
   \tablewidth{0pt}
   \centering
   \tablecaption{Best-Fit MHR00 Model Parameters (Isothermal)}
   \tablehead{ \colhead{QSO} & 
               \colhead{$\langle z_{\rm abs} \rangle$\tablenotemark{a}} &
               \colhead{$N_{\rm bin}$\tablenotemark{b}} & 
               \multicolumn{2}{c}{Continuum and zero point fixed} & &
               \multicolumn{4}{c}{Continuum and zero point allowed to vary} \\ 
               \cline{4-5} \cline{7-10}
                &  &  & \colhead{\G\tablenotemark{c}} & \colhead{$\chi^2_{r}$} &
                & \colhead{\G\tablenotemark{c}} & 
               \colhead{Cont.\tablenotemark{d}} & 
               \colhead{zero point\tablenotemark{e}} & \colhead{$\chi^2_{r}$}
   }
   \startdata
   SDSS~J1148$+$5251 & 5.614 & 49 & 0.14 &  1.32 & & 0.12 & 1.188 &  0.005   & 0.67 \\
   SDSS~J0818$+$1722 & 5.590 & 83 & 0.14 &  5.21 & & 0.11 & 2.427 &  0.000   & 3.08 \\
   SDSS~J1623$+$3112 & 5.522 & 76 & 0.15 &  0.93 & & 0.14 & 0.983 &  0.005   & 0.89 \\
   SDSS~J1048$+$4637 & 5.516 & 88 & 0.22 &  1.93 & & 0.20 & 1.195 & -0.006   & 1.38 \\
   SDSS~J1030$+$0524 & 5.514 & 83 & 0.21 &  1.30 & & 0.17 & 1.322 &  0.006   & 0.89 \\
   SDSS~J0002$+$2550 & 5.465 & 68 & 0.13 &  1.34 & & 0.12 & 1.071 &  0.009   & 1.10 \\
   SDSS~J0836$+$0054 & 5.455 & 55 & 0.20 &  1.00 & & 0.18 & 1.198 &  0.002   & 0.64 \\
   SDSS~J0818$+$1722 & 5.221 & 73 & 0.23 &  1.53 & & 0.20 & 1.175 &  0.006   & 0.78 \\
   SDSS~J0002$+$2550 & 5.076 & 75 & 0.17 &  2.40 & & 0.14 & 1.125 &  0.016   & 1.47 \\
   SDSS~J0836$+$0054 & 5.067 & 56 & 0.16 &  0.84 & & 0.16 & 1.054 &  0.003   & 0.70 \\
   SDSS~J0231$-$0728 & 5.043 & 78 & 0.30 &  1.43 & & 0.25 & 1.153 &  0.012   & 0.69 \\
   SDSS~J0915$+$4244 & 4.849 & 71 & 0.19 &  3.35 & & 0.16 & 1.304 &  0.007   & 0.97 \\
   SDSS~J1204$-$0021 & 4.747 & 75 & 0.40 &  7.42 & & 0.27 & 1.210 &  0.031   & 1.05 \\
   SDSS~J0231$-$0728 & 4.730 & 77 & 0.27 &  1.22 & & 0.25 & 1.039 &  0.012   & 0.86 \\
   SDSS~J2225$-$0014 & 4.513 & 80 & 0.30 &  3.70 & & 0.24 & 1.193 &  0.010   & 0.82 \\
   SDSS~J0915$+$4244 & 4.509 & 77 & 0.33 &  3.17 & & 0.26 & 1.137 &  0.017   & 0.75 \\
   SDSS~J1204$-$0021 & 4.428 & 79 & 0.36 &  6.64 & & 0.28 & 1.194 &  0.005   & 1.60 \\
      BRI2237$-$0607 & 4.254 & 84 & 0.64 &  4.94 & & 0.50 & 1.094 & -0.005   & 1.62 \\
   SDSS~J2225$-$0014 & 4.234 & 86 & 0.28 &  5.13 & & 0.21 & 1.170 & -0.002   & 1.29 \\
        Q0246$+$1750 & 4.123 & 65 & 0.40 &  2.55 & & 0.34 & 1.069 &  0.009   & 1.11 \\
      BRI1202$-$0725 & 4.074 & 87 & 0.29 & 10.08 & & 0.23 & 1.182 &  0.011   & 2.21 \\
        Q0246$+$1750 & 3.851 & 78 & 0.55 &  5.26 & & 0.39 & 1.073 & -0.005   & 2.11 \\
        Q1055$+$4611 & 3.846 & 62 & 0.24 &  4.68 & & 0.22 & 1.091 &  0.002   & 1.60 \\
         Q0000$-$263 & 3.833 & 70 & 0.34 &  5.50 & & 0.28 & 1.097 &  0.010   & 1.56 \\
        Q1645$+$5520 & 3.798 & 60 & 0.33 &  5.49 & & 0.31 & 1.076 &  0.002   & 1.53 \\
      BRI0241$-$0146 & 3.779 & 67 & 0.40 &  6.43 & & 0.29 & 1.076 &  0.019   & 1.86 \\
        Q0827$+$5255 & 3.623 & 55 & 0.31 &  4.93 & & 0.32 & 1.059 &  0.000   & 1.66 \\
        Q1055$+$4611 & 3.591 & 63 & 0.51 &  7.66 & & 0.41 & 1.052 &  0.004   & 1.73 \\
         Q0000$-$263 & 3.574 & 73 & 0.44 &  5.32 & & 0.32 & 1.048 &  0.002   & 3.33 \\
        Q1645$+$5520 & 3.543 & 61 & 0.33 &  4.27 & & 0.28 & 1.055 &  0.000   & 1.17 \\
      BRI0241$-$0146 & 3.523 & 74 & 0.32 &  4.96 & & 0.26 & 1.064 &  0.000   & 1.92 \\
        Q0827$+$5255 & 3.389 & 61 & 0.33 &  5.88 & & 0.27 & 1.060 &  0.004   & 1.17 \\
        Q0055$-$2659 & 3.381 & 67 & 0.69 &  3.02 & & 0.51 & 1.030 &  0.010   & 1.11 \\
       Q1422$+$2309A & 3.358 & 58 & 0.56 &  4.41 & & 0.42 & 1.037 &  0.008   & 1.16 \\
        Q0055$-$2659 & 3.149 & 73 & 0.54 &  4.75 & & 0.44 & 1.014 &  0.013   & 4.44 \\
       Q1422$+$2309A & 3.126 & 57 & 0.44 &  2.99 & & 0.37 & 1.025 &  0.004   & 0.89 \\
        Q0956$+$1217 & 3.061 & 64 & 0.37 &  4.47 & & 0.29 & 1.035 &  0.001   & 1.50 \\
        Q0930$+$2858 & 2.955 & 74 & 0.49 &  1.24 & & 0.44 & 1.011 &  \nodata & 1.02 \\
        Q0636$+$6801 & 2.931 & 56 & 0.50 &  3.02 & & 0.39 & 1.017 &  \nodata & 1.30 \\
          Q0642$+$44 & 2.927 & 76 & 0.38 &  2.12 & & 0.29 & 1.030 &  \nodata & 0.95 \\
        Q1140$+$3508 & 2.916 & 60 & 0.58 &  4.64 & & 0.39 & 1.023 &  \nodata & 2.05 \\
        Q0449$-$1326 & 2.860 & 65 & 0.38 &  1.60 & & 0.30 & 1.024 &  \nodata & 0.54 \\
        Q0956$+$1217 & 2.845 & 66 & 0.62 &  3.04 & & 0.45 & 1.018 &  \nodata & 1.75 \\
        Q0940$-$1050 & 2.844 & 61 & 0.39 &  3.14 & & 0.30 & 1.019 &  \nodata & 1.81 \\
        Q2231$-$0015 & 2.780 & 64 & 0.37 &  3.12 & & 0.30 & 1.021 &  \nodata & 1.92 \\
       HS0741$+$4741 & 2.772 & 60 & 0.49 &  4.59 & & 0.35 & 1.024 &  \nodata & 1.45 \\
       HS1011$+$4315 & 2.766 & 63 & 0.44 &  2.46 & & 0.32 & 1.019 &  \nodata & 1.13 \\
        Q0216$+$0803 & 2.748 & 86 & 0.21 &  1.30 & & 0.19 & 1.010 &  \nodata & 1.23 \\
        Q1437$+$3007 & 2.746 & 65 & 0.49 &  1.14 & & 0.45 & 1.005 &  \nodata & 1.03 \\
        Q0636$+$6801 & 2.720 & 56 & 0.35 &  2.75 & & 0.27 & 1.017 &  \nodata & 1.22 \\
        Q1140$+$3508 & 2.708 & 61 & 0.56 &  3.12 & & 0.40 & 1.014 &  \nodata & 1.66 \\
        Q1511$+$0907 & 2.658 & 70 & 0.42 &  1.74 & & 0.35 & 1.011 &  \nodata & 1.34 \\
        Q0449$-$1326 & 2.654 & 69 & 0.39 &  1.18 & & 0.45 & 0.990 &  \nodata & 0.91 \\
        Q1132$+$2243 & 2.652 & 71 & 0.39 &  2.72 & & 0.27 & 1.022 &  \nodata & 1.55 \\
       HS0119$+$1432 & 2.643 & 58 & 0.54 &  1.34 & & 0.48 & 1.005 &  \nodata & 1.15 \\
        Q0940$-$1050 & 2.639 & 66 & 0.38 &  3.85 & & 0.24 & 1.029 &  \nodata & 1.80 \\
       HS1946$+$7658 & 2.627 & 59 & 0.36 &  2.45 & & 0.28 & 1.015 &  \nodata & 1.10 \\
        Q1549$+$1919 & 2.613 & 53 & 0.50 &  3.14 & & 0.43 & 1.011 &  \nodata & 1.17 \\
         Q0528$-$250 & 2.595 & 65 & 0.25 &  5.44 & & 0.16 & 1.022 &  \nodata & 4.54 \\
        Q2231$-$0015 & 2.578 & 74 & 0.29 &  1.56 & & 0.26 & 1.007 &  \nodata & 1.52 \\
        Q0216$+$0803 & 2.574 & 89 & 0.18 &  1.32 & & 0.19 & 0.991 &  \nodata & 1.31 \\
        Q1437$+$3007 & 2.547 & 73 & 0.45 &  1.44 & & 0.41 & 1.005 &  \nodata & 1.43 \\
         Q1107$+$487 & 2.546 & 72 & 0.55 &  1.05 & & 0.49 & 1.005 &  \nodata & 0.96 \\
        Q1244$+$3133 & 2.541 & 81 & 0.22 &  2.34 & & 0.20 & 1.011 &  \nodata & 2.12 \\
       HS1700$+$6416 & 2.525 & 54 & 0.57 &  2.41 & & 0.44 & 1.007 &  \nodata & 1.57 \\
        Q1511$+$0907 & 2.464 & 76 & 0.31 &  1.03 & & 0.30 & 1.002 &  \nodata & 1.02 \\
        Q1132$+$2243 & 2.456 & 74 & 0.43 &  1.28 & & 0.48 & 0.994 &  \nodata & 1.21 \\
       HS0119$+$1432 & 2.452 & 60 & 0.32 &  1.46 & & 0.29 & 1.005 &  \nodata & 1.36 \\
        Q1009$+$2956 & 2.436 & 56 & 0.48 &  2.05 & & 0.43 & 1.004 &  \nodata & 1.83 \\
        Q1549$+$1919 & 2.419 & 55 & 0.68 &  3.77 & & 0.48 & 1.008 &  \nodata & 2.60 \\
         Q0528$-$250 & 2.398 & 72 & 0.30 &  1.10 & & 0.34 & 0.993 &  \nodata & 0.99 \\
        Q2344$+$1228 & 2.374 & 79 & 0.29 &  2.15 & & 0.28 & 1.002 &  \nodata & 2.17 \\
        Q1358$+$1134 & 2.370 & 86 & 0.10 &  3.11 & & 0.16 & 0.942 &  \nodata & 0.85 \\
        Q2206$-$199N & 2.356 & 60 & 0.37 &  1.09 & & 0.33 & 1.004 &  \nodata & 1.02 \\
       HS1700$+$6416 & 2.339 & 55 & 0.41 &  1.78 & & 0.41 & 1.000 &  \nodata & 1.89 \\
        Q1623$+$2653 & 2.323 & 67 & 0.41 &  1.69 & & 0.36 & 1.004 &  \nodata & 1.55 \\
        Q1442$+$2931 & 2.264 & 60 & 0.53 &  0.99 & & 0.62 & 0.996 &  \nodata & 0.77 \\
        Q1009$+$2956 & 2.252 & 59 & 0.36 &  0.67 & & 0.33 & 1.003 &  \nodata & 0.62 \\
        Q2343$+$1232 & 2.190 & 81 & 0.28 &  1.14 & & 0.23 & 1.010 &  \nodata & 1.05 \\
        Q2206$-$199N & 2.188 & 61 & 0.33 &  0.97 & & 0.31 & 1.002 &  \nodata & 0.99 \\
        Q1623$+$2653 & 2.146 & 77 & 0.29 &  1.30 & & 0.35 & 0.990 &  \nodata & 1.07 \\
        Q0841$+$1256 & 2.127 & 81 & 0.14 &  2.42 & & 0.22 & 0.970 &  \nodata & 1.11 \\
         Q0237$-$233 & 2.050 & 67 & 0.13 &  5.61 & & 0.28 & 0.967 &  \nodata & 1.30 \\
        Q1225$+$3145 & 2.016 & 62 & 0.32 &  2.00 & & 0.30 & 1.002 &  \nodata & 2.01 \\
         Q0421$+$019 & 1.870 & 70 & 0.49 &  5.17 & & 0.95 & 0.979 &  \nodata & 2.97 \\
        Q1225$+$3145 & 1.857 & 63 & 0.25 &  1.71 & & 0.24 & 1.001 &  \nodata & 1.70 \\
        Q0119$-$0437 & 1.807 & 83 & 0.30 &  4.33 & & 0.74 & 0.963 &  \nodata & 1.93 \\
        Q0058$+$0155 & 1.797 & 79 & 0.29 &  3.75 & & 0.64 & 0.966 &  \nodata & 1.96
   \enddata
   \tablenotetext{a}{Mean absorption redshift.}
   \tablenotetext{b}{Number of flux bins over which fit was performed.}
   \tablenotetext{c}{\hi\ ionization rate, in units of $10^{-1}$~s$^{-1}$.}
   \tablenotetext{d}{Factor by which to multiply the continuum
     in order for the model to produce the best fit.}
   \tablenotetext{e}{Flux zero point that would allow the model to
     produce the best fit.}
   \label{tab:mhr00_params_isothermal}
\end{deluxetable}

\clearpage

\begin{deluxetable}{lccccccccccc}
   \tabletypesize{\scriptsize}
   \tablewidth{0pt}
   \centering
   \tablecaption{Best-Fit Lognormal Parameters}
   \tablehead{ \colhead{QSO} & 
     \colhead{$\langle z_{\rm abs} \rangle$\tablenotemark{a}}
     & \colhead{$N_{\rm bin}$\tablenotemark{b}} & 
     \multicolumn{3}{c}{Continuum and zero point fixed} & &
     \multicolumn{5}{c}{Continuum and zero point allowed to vary} \\ 
     \cline{4-6} \cline{8-12}
      &  &  & \colhead{$\mu$\tablenotemark{c}} & 
     \colhead{$\sigma$\tablenotemark{d}} & \colhead{\rchisq} &
      & \colhead{$\mu$\tablenotemark{c}} & 
     \colhead{$\sigma$\tablenotemark{d}} & \colhead{Cont.\tablenotemark{e}} & 
     \colhead{Zero pt.\tablenotemark{f}} & \colhead{$\chi^2_{r}$}
   }
   \startdata
   SDSS~J1148$+$5251 & 5.614 & 49 &  1.81 & 0.86 & 1.10 & &  2.21 & 1.24 & 0.774 &  0.009   & 0.49 \\
   SDSS~J0818$+$1722 & 5.590 & 83 &  2.71 & 1.83 & 1.98 & &  2.59 & 1.63 & 1.161 &  0.002   & 1.91 \\
   SDSS~J1623$+$3112 & 5.522 & 76 &  1.58 & 0.80 & 0.83 & &  1.56 & 0.75 & 1.088 & -0.001   & 0.87 \\
   SDSS~J1048$+$4637 & 5.516 & 88 &  1.58 & 1.19 & 1.42 & &  1.40 & 0.83 & 1.427 & -0.010   & 0.99 \\
   SDSS~J1030$+$0524 & 5.514 & 83 &  1.53 & 1.07 & 0.97 & &  1.63 & 1.16 & 0.983 &  0.006   & 0.92 \\
   SDSS~J0002$+$2550 & 5.465 & 68 &  1.61 & 0.79 & 1.21 & &  1.82 & 0.94 & 0.940 &  0.009   & 1.14 \\
   SDSS~J0836$+$0054 & 5.455 & 55 &  1.54 & 1.10 & 0.64 & &  1.66 & 1.36 & 0.866 &  0.004   & 0.47 \\
   SDSS~J0818$+$1722 & 5.221 & 73 &  1.19 & 1.15 & 0.88 & &  1.29 & 1.31 & 0.937 &  0.007   & 0.71 \\
   SDSS~J0002$+$2550 & 5.076 & 75 &  1.19 & 0.95 & 1.90 & &  1.42 & 1.25 & 0.884 &  0.019   & 1.43 \\
   SDSS~J0836$+$0054 & 5.067 & 56 &  1.29 & 1.00 & 0.72 & &  1.29 & 1.10 & 0.922 &  0.000   & 0.59 \\
   SDSS~J0231$-$0728 & 5.043 & 78 &  0.80 & 1.23 & 0.91 & &  0.89 & 1.35 & 0.974 &  0.012   & 0.72 \\
   SDSS~J0915$+$4244 & 4.849 & 71 &  1.12 & 1.49 & 1.13 & &  1.28 & 1.61 & 1.004 &  0.011   & 0.48 \\
   SDSS~J1204$-$0021 & 4.747 & 75 &  0.26 & 1.36 & 4.23 & &  0.58 & 1.49 & 1.052 &  0.033   & 0.99 \\
   SDSS~J0231$-$0728 & 4.730 & 77 &  0.63 & 1.17 & 1.22 & &  0.64 & 1.38 & 0.916 &  0.009   & 0.96 \\
   SDSS~J2225$-$0014 & 4.513 & 80 &  0.40 & 1.65 & 0.95 & &  0.49 & 1.63 & 1.028 &  0.012   & 0.63 \\
   SDSS~J0915$+$4244 & 4.509 & 77 &  0.26 & 1.40 & 1.60 & &  0.40 & 1.43 & 1.036 &  0.014   & 0.84 \\
   SDSS~J1204$-$0021 & 4.428 & 79 &  0.23 & 1.91 & 1.28 & &  0.31 & 1.87 & 1.026 &  0.009   & 0.98 \\
      BRI2237$-$0607 & 4.254 & 84 & -0.60 & 1.98 & 1.67 & & -0.60 & 1.92 & 1.004 &  0.008   & 1.64 \\
   SDSS~J2225$-$0014 & 4.234 & 86 &  0.34 & 1.82 & 0.71 & &  0.35 & 1.88 & 0.991 &  0.004   & 0.69 \\
        Q0246$+$1750 & 4.123 & 65 & -0.35 & 1.56 & 1.54 & & -0.32 & 1.51 & 1.016 &  0.002   & 1.49 \\
      BRI1202$-$0725 & 4.074 & 87 & -0.08 & 2.04 & 2.19 & &  0.05 & 1.98 & 1.032 &  0.018   & 1.24 \\
        Q0246$+$1750 & 3.851 & 78 & -0.74 & 2.16 & 1.36 & & -0.79 & 2.26 & 0.987 &  0.000   & 1.26 \\
        Q1055$+$4611 & 3.846 & 62 & -0.15 & 2.02 & 0.89 & & -0.12 & 1.96 & 1.019 &  0.003   & 0.58 \\
         Q0000$-$263 & 3.833 & 70 & -0.59 & 1.93 & 1.94 & & -0.45 & 1.94 & 1.018 &  0.010   & 1.19 \\
        Q1645$+$5520 & 3.798 & 60 & -0.54 & 2.06 & 1.15 & & -0.53 & 1.96 & 1.022 &  0.002   & 0.76 \\
      BRI0241$-$0146 & 3.779 & 67 & -0.78 & 1.85 & 2.39 & & -0.61 & 1.99 & 1.004 &  0.020   & 1.07 \\
        Q0827$+$5255 & 3.623 & 55 & -0.58 & 2.30 & 0.80 & & -0.65 & 2.20 & 1.015 &  0.000   & 0.60 \\
        Q1055$+$4611 & 3.591 & 63 & -1.31 & 2.15 & 1.23 & & -1.24 & 2.08 & 1.008 &  0.003   & 0.96 \\
         Q0000$-$263 & 3.574 & 73 & -0.75 & 2.25 & 1.57 & & -0.82 & 2.53 & 0.981 &  0.008   & 1.05 \\
        Q1645$+$5520 & 3.543 & 61 & -0.85 & 2.07 & 0.76 & & -0.85 & 1.98 & 1.009 &  0.002   & 0.70 \\
      BRI0241$-$0146 & 3.523 & 74 & -0.82 & 2.07 & 1.76 & & -0.83 & 2.10 & 0.997 &  0.001   & 1.80 \\
        Q0827$+$5255 & 3.389 & 61 & -1.22 & 1.97 & 1.58 & & -1.13 & 1.87 & 1.018 &  0.004   & 0.88 \\
        Q0055$-$2659 & 3.381 & 67 & -1.88 & 2.04 & 1.17 & & -1.92 & 2.18 & 0.994 &  0.008   & 1.05 \\
       Q1422$+$2309A & 3.358 & 58 & -1.79 & 2.03 & 1.56 & & -1.62 & 2.16 & 1.005 &  0.007   & 0.82 \\
        Q0055$-$2659 & 3.149 & 73 & -1.67 & 2.48 & 3.41 & & -2.16 & 3.32 & 0.970 &  0.011   & 1.15 \\
       Q1422$+$2309A & 3.126 & 57 & -1.95 & 1.98 & 1.05 & & -1.89 & 1.87 & 1.005 &  0.002   & 0.97 \\
        Q0956$+$1217 & 3.061 & 64 & -1.85 & 2.16 & 1.63 & & -1.83 & 2.07 & 1.004 &  0.002   & 1.67 \\
        Q0930$+$2858 & 2.955 & 74 & -2.28 & 1.99 & 1.05 & & -2.44 & 2.15 & 0.988 &  \nodata & 0.89 \\
        Q0636$+$6801 & 2.931 & 56 & -2.23 & 2.13 & 0.78 & & -2.25 & 2.17 & 0.999 &  \nodata & 0.79 \\
          Q0642$+$44 & 2.927 & 76 & -2.00 & 2.20 & 0.72 & & -2.10 & 2.31 & 0.992 &  \nodata & 0.67 \\
        Q1140$+$3508 & 2.916 & 60 & -2.32 & 2.40 & 0.89 & & -2.35 & 2.43 & 0.998 &  \nodata & 0.93 \\
        Q0449$-$1326 & 2.860 & 65 & -2.16 & 2.07 & 0.80 & & -2.16 & 2.06 & 1.001 &  \nodata & 0.83 \\
        Q0956$+$1217 & 2.845 & 66 & -2.65 & 2.54 & 1.15 & & -2.86 & 2.81 & 0.991 &  \nodata & 0.78 \\
        Q0940$-$1050 & 2.844 & 61 & -2.07 & 2.28 & 1.31 & & -2.22 & 2.54 & 0.990 &  \nodata & 0.93 \\
        Q2231$-$0015 & 2.780 & 64 & -2.25 & 2.03 & 1.48 & & -2.20 & 2.00 & 1.003 &  \nodata & 1.50 \\
       HS0741$+$4741 & 2.772 & 60 & -2.57 & 2.36 & 1.14 & & -2.46 & 2.22 & 1.006 &  \nodata & 0.97 \\
       HS1011$+$4315 & 2.766 & 63 & -2.39 & 2.25 & 0.61 & & -2.45 & 2.33 & 0.997 &  \nodata & 0.58 \\
        Q0216$+$0803 & 2.748 & 86 & -1.65 & 1.91 & 1.22 & & -1.99 & 2.22 & 0.971 &  \nodata & 0.96 \\
        Q1437$+$3007 & 2.746 & 65 & -2.65 & 1.96 & 1.75 & & -2.89 & 2.29 & 0.988 &  \nodata & 1.24 \\
        Q0636$+$6801 & 2.720 & 56 & -2.26 & 2.11 & 0.90 & & -2.28 & 2.14 & 0.999 &  \nodata & 0.91 \\
        Q1140$+$3508 & 2.708 & 61 & -2.65 & 2.41 & 0.74 & & -2.81 & 2.60 & 0.994 &  \nodata & 0.52 \\
        Q1511$+$0907 & 2.658 & 70 & -2.62 & 2.21 & 1.40 & & -2.93 & 2.58 & 0.987 &  \nodata & 0.95 \\
        Q0449$-$1326 & 2.654 & 69 & -2.52 & 1.60 & 2.20 & & -3.13 & 2.32 & 0.974 &  \nodata & 1.10 \\
        Q1132$+$2243 & 2.652 & 71 & -2.51 & 2.43 & 0.92 & & -2.66 & 2.59 & 0.993 &  \nodata & 0.80 \\
       HS0119$+$1432 & 2.643 & 58 & -2.82 & 2.08 & 1.30 & & -3.13 & 2.40 & 0.991 &  \nodata & 0.73 \\
        Q0940$-$1050 & 2.639 & 66 & -2.35 & 2.62 & 0.66 & & -2.43 & 2.73 & 0.996 &  \nodata & 0.63 \\
       HS1946$+$7658 & 2.627 & 59 & -2.46 & 2.12 & 0.79 & & -2.51 & 2.17 & 0.998 &  \nodata & 0.80 \\
        Q1549$+$1919 & 2.613 & 53 & -3.02 & 2.14 & 1.10 & & -2.92 & 2.01 & 1.003 &  \nodata & 0.97 \\
         Q0528$-$250 & 2.595 & 65 & -1.68 & 2.41 & 2.35 & & -2.11 & 2.83 & 0.982 &  \nodata & 1.50 \\
        Q2231$-$0015 & 2.578 & 74 & -2.37 & 2.26 & 2.00 & & -2.87 & 2.85 & 0.975 &  \nodata & 0.97 \\
        Q0216$+$0803 & 2.574 & 89 & -1.90 & 1.97 & 2.02 & & -2.83 & 2.98 & 0.939 &  \nodata & 0.95 \\
        Q1437$+$3007 & 2.547 & 73 & -3.00 & 2.34 & 1.94 & & -3.49 & 3.01 & 0.982 &  \nodata & 0.98 \\
         Q1107$+$487 & 2.546 & 72 & -3.17 & 2.27 & 1.17 & & -3.55 & 2.61 & 0.989 &  \nodata & 0.72 \\
        Q1244$+$3133 & 2.541 & 81 & -2.16 & 2.46 & 2.13 & & -2.78 & 3.15 & 0.966 &  \nodata & 0.74 \\
       HS1700$+$6416 & 2.525 & 54 & -3.08 & 2.38 & 1.18 & & -3.31 & 2.81 & 0.995 &  \nodata & 0.67 \\
        Q1511$+$0907 & 2.464 & 76 & -2.66 & 1.97 & 1.03 & & -2.88 & 2.17 & 0.989 &  \nodata & 0.85 \\
        Q1132$+$2243 & 2.456 & 74 & -3.05 & 1.99 & 1.85 & & -3.76 & 2.60 & 0.979 &  \nodata & 1.02 \\
       HS0119$+$1432 & 2.452 & 60 & -2.66 & 2.22 & 1.94 & & -3.10 & 2.71 & 0.985 &  \nodata & 0.64 \\
        Q1009$+$2956 & 2.436 & 56 & -3.12 & 2.14 & 2.13 & & -3.50 & 2.64 & 0.992 &  \nodata & 1.34 \\
        Q1549$+$1919 & 2.419 & 55 & -3.46 & 2.84 & 1.38 & & -3.63 & 3.16 & 0.996 &  \nodata & 1.09 \\
         Q0528$-$250 & 2.398 & 72 & -2.79 & 2.02 & 2.29 & & -3.53 & 2.68 & 0.974 &  \nodata & 0.85 \\
        Q2344$+$1228 & 2.374 & 79 & -2.91 & 2.60 & 2.42 & & -3.76 & 3.44 & 0.973 &  \nodata & 0.89 \\
        Q1358$+$1134 & 2.370 & 86 & -1.63 & 1.47 & 4.00 & & -2.82 & 2.67 & 0.908 &  \nodata & 0.73 \\
        Q2206$-$199N & 2.356 & 60 & -3.07 & 2.15 & 1.44 & & -3.48 & 2.71 & 0.990 &  \nodata & 0.80 \\
       HS1700$+$6416 & 2.339 & 55 & -3.16 & 1.97 & 3.81 & & -3.75 & 2.97 & 0.989 &  \nodata & 1.39 \\
        Q1623$+$2653 & 2.323 & 67 & -3.24 & 2.26 & 1.41 & & -3.53 & 2.48 & 0.992 &  \nodata & 1.12 \\
        Q1442$+$2931 & 2.264 & 60 & -3.58 & 1.83 & 1.75 & & -4.18 & 2.39 & 0.990 &  \nodata & 0.93 \\
        Q1009$+$2956 & 2.252 & 59 & -3.30 & 2.10 & 1.08 & & -3.60 & 2.50 & 0.993 &  \nodata & 0.74 \\
        Q2343$+$1232 & 2.190 & 81 & -3.37 & 2.74 & 0.93 & & -3.86 & 3.16 & 0.985 &  \nodata & 0.65 \\
        Q2206$-$199N & 2.188 & 61 & -3.34 & 2.17 & 1.51 & & -3.82 & 2.69 & 0.989 &  \nodata & 0.98 \\
        Q1623$+$2653 & 2.146 & 77 & -3.18 & 1.89 & 1.64 & & -3.89 & 2.49 & 0.981 &  \nodata & 1.19 \\
        Q0841$+$1256 & 2.127 & 81 & -2.43 & 1.72 & 2.92 & & -3.58 & 2.69 & 0.953 &  \nodata & 0.95 \\
         Q0237$-$233 & 2.050 & 67 & -2.49 & 1.35 & 3.61 & & -3.59 & 2.31 & 0.961 &  \nodata & 0.88 \\
        Q1225$+$3145 & 2.016 & 62 & -3.62 & 2.42 & 1.64 & & -3.98 & 2.70 & 0.994 &  \nodata & 1.19 \\
         Q0421$+$019 & 1.870 & 70 & -3.81 & 1.92 & 1.27 & & -4.51 & 2.34 & 0.986 &  \nodata & 0.76 \\
        Q1225$+$3145 & 1.857 & 63 & -3.91 & 2.56 & 2.45 & & -4.80 & 3.56 & 0.989 &  \nodata & 1.37 \\
        Q0119$-$0437 & 1.807 & 83 & -3.36 & 1.87 & 1.38 & & -4.42 & 2.49 & 0.973 &  \nodata & 0.83 \\
        Q0058$+$0155 & 1.797 & 79 & -3.39 & 1.87 & 0.87 & & -4.13 & 2.32 & 0.979 &  \nodata & 0.59
   \enddata
   \tablenotetext{a}{Mean absorption redshift.}
   \tablenotetext{b}{Number of flux bins over which fit was performed.}
   \tablenotetext{c}{${\rm L}$ognormal parameter $\mu = \langle
     \ln{\tau} \rangle$.}
   \tablenotetext{d}{${\rm L}$ognormal parameter $\sigma = $
     std\,dev\,$(\ln{\tau})$.}  
   \tablenotetext{e}{Factor by which to
     multiply the continuum in order for the model to produce the best
     fit.}  
   \tablenotetext{f}{Flux zero point that would allow the
     model to produce the best fit.}
   \label{tab:lognormal_params}
\end{deluxetable}

\clearpage

\begin{deluxetable}{lccccccccccc}
   \tabletypesize{\scriptsize}
   \tablewidth{0pt}
   \centering
   \tablecaption{Best-Fit MHR00 Model Parameters (Non-Isothermal)}
   \tablehead{ \colhead{QSO} & 
     \colhead{$\langle z_{\rm abs} \rangle$\tablenotemark{a}}
     & \colhead{$N_{\rm bin}$\tablenotemark{b}} & 
     \multicolumn{3}{c}{Continuum and zero point fixed} & &
     \multicolumn{5}{c}{Continuum and zero point allowed to vary} \\ 
     \cline{4-6} \cline{8-12}
      &  &  & \colhead{\G\tablenotemark{c}} & 
     \colhead{$\alpha$\tablenotemark{d}} & \colhead{\rchisq} &
      & \colhead{\G\tablenotemark{c}} & 
     \colhead{$\alpha$\tablenotemark{d}} & \colhead{Cont.\tablenotemark{e}} & 
     \colhead{Zero pt.\tablenotemark{f}} & \colhead{$\chi^2_{r}$}
   }
   \startdata
   SDSS~J1148$+$5251  & 5.614 & 49 & 0.13 & -0.06 & 1.30 & & 0.04 & -1.06 & 0.764 &  0.011   & 0.45 \\
   SDSS~J0818$+$1722  & 5.590 & 83 & 0.01 & -2.21 & 2.16 & & 0.02 & -1.77 & 1.195 &  0.003   & 1.88 \\
   SDSS~J1623$+$3112  & 5.522 & 76 & 0.20 &  0.25 & 0.86 & & 0.17 &  0.18 & 1.050 &  0.003   & 0.89 \\
   SDSS~J1048$+$4637  & 5.516 & 88 & 0.13 & -0.49 & 1.51 & & 0.19 &  0.05 & 1.431 & -0.002   & 1.21 \\
   SDSS~J1030$+$0524  & 5.514 & 83 & 0.16 & -0.23 & 1.17 & & 0.13 & -0.34 & 1.102 &  0.008   & 0.93 \\
   SDSS~J0002$+$2550  & 5.465 & 68 & 0.18 &  0.23 & 1.24 & & 0.10 & -0.15 & 0.986 &  0.010   & 1.11 \\
   SDSS~J0836$+$0054  & 5.455 & 55 & 0.14 & -0.37 & 0.61 & & 0.11 & -0.54 & 0.981 &  0.005   & 0.47 \\
   SDSS~J0818$+$1722  & 5.221 & 73 & 0.17 & -0.30 & 1.21 & & 0.16 & -0.26 & 1.081 &  0.008   & 0.77 \\
   SDSS~J0002$+$2550  & 5.076 & 75 & 0.18 &  0.06 & 2.44 & & 0.11 & -0.31 & 0.986 &  0.021   & 1.42 \\
   SDSS~J0836$+$0054  & 5.067 & 56 & 0.16 &  0.01 & 0.84 & & 0.18 &  0.18 & 1.136 &  0.001   & 0.69 \\
   SDSS~J0231$-$0728  & 5.043 & 78 & 0.25 & -0.19 & 1.27 & & 0.23 & -0.12 & 1.119 &  0.014   & 0.70 \\
   SDSS~J0915$+$4244  & 4.849 & 71 & 0.10 & -0.74 & 1.90 & & 0.09 & -0.67 & 1.100 &  0.012   & 0.49 \\
   SDSS~J1204$-$0021  & 4.747 & 75 & 0.37 & -0.18 & 7.44 & & 0.24 & -0.17 & 1.166 &  0.033   & 1.04 \\
   SDSS~J0231$-$0728  & 4.730 & 77 & 0.27 & -0.01 & 1.27 & & 0.23 & -0.11 & 1.006 &  0.013   & 0.90 \\
   SDSS~J2225$-$0014  & 4.513 & 80 & 0.19 & -0.61 & 1.82 & & 0.20 & -0.31 & 1.120 &  0.015   & 0.65 \\
   SDSS~J0915$+$4244  & 4.509 & 77 & 0.30 & -0.11 & 3.14 & & 0.26 & -0.01 & 1.126 &  0.018   & 0.76 \\
   SDSS~J1204$-$0021  & 4.428 & 79 & 0.17 & -0.92 & 1.95 & & 0.18 & -0.63 & 1.086 &  0.012   & 0.80 \\
      BRI2237$-$0607  & 4.254 & 84 & 0.35 & -0.78 & 1.99 & & 0.43 & -0.31 & 1.056 & -0.002   & 1.45 \\
   SDSS~J2225$-$0014  & 4.234 & 86 & 0.13 & -0.84 & 0.97 & & 0.13 & -0.68 & 1.043 &  0.009   & 0.72 \\
        Q0246$+$1750  & 4.123 & 65 & 0.36 & -0.12 & 2.44 & & 0.38 &  0.16 & 1.090 &  0.007   & 1.06 \\
      BRI1202$-$0725  & 4.074 & 87 & 0.15 & -1.03 & 3.94 & & 0.16 & -0.61 & 1.090 &  0.022   & 1.40 \\
        Q0246$+$1750  & 3.851 & 78 & 0.23 & -0.88 & 1.12 & & 0.24 & -0.71 & 1.021 &  0.005   & 1.00 \\
        Q1055$+$4611  & 3.846 & 62 & 0.11 & -0.87 & 1.91 & & 0.13 & -0.62 & 1.049 &  0.004   & 0.72 \\
        Q0000$-$263   & 3.833 & 70 & 0.22 & -0.62 & 3.52 & & 0.21 & -0.40 & 1.063 &  0.012   & 1.22 \\
        Q1645$+$5520  & 3.798 & 60 & 0.16 & -0.87 & 2.16 & & 0.20 & -0.53 & 1.050 &  0.003   & 0.81 \\
      BRI0241$-$0146  & 3.779 & 67 & 0.30 & -0.48 & 4.82 & & 0.22 & -0.43 & 1.041 &  0.023   & 1.32 \\
        Q0827$+$5255  & 3.623 & 55 & 0.09 & -1.19 & 1.08 & & 0.13 & -0.83 & 1.034 &  0.001   & 0.43 \\
        Q1055$+$4611  & 3.591 & 63 & 0.27 & -0.84 & 3.18 & & 0.29 & -0.44 & 1.033 &  0.005   & 1.01 \\
        Q0000$-$263   & 3.574 & 73 & 0.15 & -0.99 & 1.25 & & 0.13 & -1.11 & 1.000 &  0.010   & 0.94 \\
        Q1645$+$5520  & 3.543 & 61 & 0.14 & -0.83 & 1.30 & & 0.18 & -0.49 & 1.032 &  0.001   & 0.53 \\
      BRI0241$-$0146  & 3.523 & 74 & 0.17 & -0.70 & 1.79 & & 0.18 & -0.46 & 1.033 &  0.005   & 1.31 \\
        Q0827$+$5255  & 3.389 & 61 & 0.22 & -0.55 & 3.70 & & 0.22 & -0.25 & 1.046 &  0.005   & 1.06 \\
        Q0055$-$2659  & 3.381 & 67 & 0.49 & -0.41 & 1.74 & & 0.43 & -0.31 & 1.017 &  0.012   & 0.86 \\
        Q1422$+$2309A & 3.358 & 58 & 0.43 & -0.38 & 3.47 & & 0.30 & -0.41 & 1.024 &  0.008   & 0.79 \\
        Q0055$-$2659  & 3.149 & 73 & 0.19 & -0.91 & 1.95 & & 0.13 & -1.65 & 0.978 &  0.013   & 1.07 \\
        Q1422$+$2309A & 3.126 & 57 & 0.30 & -0.48 & 2.13 & & 0.32 & -0.15 & 1.021 &  0.004   & 0.85 \\
        Q0956$+$1217  & 3.061 & 64 & 0.21 & -0.68 & 2.27 & & 0.24 & -0.24 & 1.026 &  0.003   & 1.45 \\
        Q0930$+$2858  & 2.955 & 74 & 0.44 & -0.14 & 1.15 & & 0.43 & -0.02 & 1.010 &  \nodata & 1.05 \\
        Q0636$+$6801  & 2.931 & 56 & 0.26 & -0.60 & 1.30 & & 0.29 & -0.35 & 1.010 &  \nodata & 0.94 \\
        Q0642$+$44    & 2.927 & 76 & 0.25 & -0.48 & 0.91 & & 0.26 & -0.27 & 1.015 &  \nodata & 0.77 \\
        Q1140$+$3508  & 2.916 & 60 & 0.27 & -0.72 & 1.82 & & 0.29 & -0.44 & 1.012 &  \nodata & 1.44 \\
        Q0449$-$1326  & 2.860 & 65 & 0.26 & -0.40 & 0.98 & & 0.29 & -0.06 & 1.021 &  \nodata & 0.56 \\
        Q0956$+$1217  & 2.845 & 66 & 0.28 & -0.78 & 0.70 & & 0.28 & -0.75 & 1.001 &  \nodata & 0.71 \\
        Q0940$-$1050  & 2.844 & 61 & 0.18 & -0.69 & 1.01 & & 0.18 & -0.61 & 1.003 &  \nodata & 1.00 \\
        Q2231$-$0015  & 2.780 & 64 & 0.31 & -0.23 & 2.80 & & 0.31 &  0.13 & 1.026 &  \nodata & 1.74 \\
       HS0741$+$4741  & 2.772 & 60 & 0.30 & -0.59 & 2.64 & & 0.32 & -0.16 & 1.020 &  \nodata & 1.41 \\
       HS1011$+$4315  & 2.766 & 63 & 0.25 & -0.55 & 1.08 & & 0.27 & -0.30 & 1.011 &  \nodata & 0.88 \\
        Q0216$+$0803  & 2.748 & 86 & 0.19 & -0.13 & 1.25 & & 0.19 & -0.10 & 1.002 &  \nodata & 1.27 \\
        Q1437$+$3007  & 2.746 & 65 & 0.42 & -0.16 & 1.02 & & 0.43 & -0.11 & 1.002 &  \nodata & 1.04 \\
        Q0636$+$6801  & 2.720 & 56 & 0.22 & -0.46 & 1.67 & & 0.24 & -0.16 & 1.013 &  \nodata & 1.16 \\
        Q1140$+$3508  & 2.708 & 61 & 0.26 & -0.69 & 0.95 & & 0.26 & -0.57 & 1.004 &  \nodata & 0.92 \\
        Q1511$+$0907  & 2.658 & 70 & 0.28 & -0.42 & 1.04 & & 0.28 & -0.46 & 0.999 &  \nodata & 1.10 \\
        Q0449$-$1326  & 2.654 & 69 & 0.45 &  0.18 & 1.03 & & 0.45 & -0.04 & 0.989 &  \nodata & 0.94 \\
        Q1132$+$2243  & 2.652 & 71 & 0.22 & -0.60 & 1.13 & & 0.22 & -0.48 & 1.006 &  \nodata & 1.10 \\
       HS0119$+$1432  & 2.643 & 58 & 0.43 & -0.24 & 1.03 & & 0.43 & -0.23 & 1.000 &  \nodata & 1.09 \\
        Q0940$-$1050  & 2.639 & 66 & 0.12 & -0.95 & 0.83 & & 0.13 & -0.75 & 1.008 &  \nodata & 0.75 \\
       HS1946$+$7658  & 2.627 & 59 & 0.25 & -0.37 & 1.58 & & 0.27 & -0.06 & 1.013 &  \nodata & 1.08 \\
        Q1549$+$1919  & 2.613 & 53 & 0.29 & -0.66 & 1.87 & & 0.38 & -0.18 & 1.009 &  \nodata & 1.14 \\
        Q0528$-$250   & 2.595 & 65 & 0.08 & -0.80 & 2.31 & & 0.08 & -1.01 & 0.991 &  \nodata & 2.26 \\
        Q2231$-$0015  & 2.578 & 74 & 0.19 & -0.38 & 1.05 & & 0.18 & -0.61 & 0.989 &  \nodata & 0.96 \\
        Q0216$+$0803  & 2.574 & 89 & 0.16 & -0.10 & 1.33 & & 0.15 & -0.81 & 0.952 &  \nodata & 0.95 \\
        Q1437$+$3007  & 2.547 & 73 & 0.29 & -0.45 & 0.93 & & 0.27 & -0.73 & 0.991 &  \nodata & 0.78 \\
        Q1107$+$487   & 2.546 & 72 & 0.45 & -0.24 & 0.81 & & 0.45 & -0.26 & 0.999 &  \nodata & 0.81 \\
        Q1244$+$3133  & 2.541 & 81 & 0.12 & -0.57 & 1.17 & & 0.11 & -1.03 & 0.976 &  \nodata & 0.80 \\
       HS1700$+$6416  & 2.525 & 54 & 0.20 & -0.86 & 0.58 & & 0.18 & -0.98 & 0.998 &  \nodata & 0.57 \\
        Q1511$+$0907  & 2.464 & 76 & 0.32 &  0.05 & 1.01 & & 0.31 &  0.16 & 1.009 &  \nodata & 1.00 \\
        Q1132$+$2243  & 2.456 & 74 & 0.44 &  0.04 & 1.30 & & 0.47 & -0.19 & 0.989 &  \nodata & 1.21 \\
       HS0119$+$1432  & 2.452 & 60 & 0.22 & -0.33 & 0.88 & & 0.22 & -0.49 & 0.995 &  \nodata & 0.81 \\
        Q1009$+$2956  & 2.436 & 56 & 0.28 & -0.47 & 1.28 & & 0.25 & -0.72 & 0.996 &  \nodata & 1.26 \\
        Q1549$+$1919  & 2.419 & 55 & 0.15 & -1.19 & 0.92 & & 0.15 & -1.24 & 0.999 &  \nodata & 0.94 \\
        Q0528$-$250   & 2.398 & 72 & 0.31 &  0.03 & 1.09 & & 0.32 & -0.25 & 0.985 &  \nodata & 0.88 \\
        Q2344$+$1228  & 2.374 & 79 & 0.18 & -0.49 & 1.48 & & 0.17 & -1.06 & 0.979 &  \nodata & 1.00 \\
        Q1358$+$1134  & 2.370 & 86 & 0.15 &  0.44 & 1.86 & & 0.15 & -0.39 & 0.922 &  \nodata & 0.74 \\
        Q2206$-$199N  & 2.356 & 60 & 0.26 & -0.33 & 0.77 & & 0.26 & -0.42 & 0.998 &  \nodata & 0.77 \\
       HS1700$+$6416  & 2.339 & 55 & 0.30 & -0.26 & 1.62 & & 0.23 & -0.73 & 0.994 &  \nodata & 1.15 \\
        Q1623$+$2653  & 2.323 & 67 & 0.37 & -0.12 & 1.56 & & 0.37 & -0.05 & 1.003 &  \nodata & 1.63 \\
        Q1442$+$2931  & 2.264 & 60 & 0.63 &  0.20 & 0.85 & & 0.63 &  0.05 & 0.996 &  \nodata & 0.82 \\
        Q1009$+$2956  & 2.252 & 59 & 0.29 & -0.21 & 0.55 & & 0.29 & -0.20 & 1.000 &  \nodata & 0.56 \\
        Q2343$+$1232  & 2.190 & 81 & 0.19 & -0.46 & 0.72 & & 0.19 & -0.55 & 0.996 &  \nodata & 0.72 \\
        Q2206$-$199N  & 2.188 & 61 & 0.29 & -0.15 & 0.91 & & 0.28 & -0.21 & 0.998 &  \nodata & 0.92 \\
        Q1623$+$2653  & 2.146 & 77 & 0.34 &  0.24 & 1.08 & & 0.35 &  0.13 & 0.994 &  \nodata & 1.08 \\
        Q0841$+$1256  & 2.127 & 81 & 0.20 &  0.39 & 1.62 & & 0.22 & -0.05 & 0.967 &  \nodata & 1.11 \\
        Q0237$-$233   & 2.050 & 67 & 0.27 &  0.68 & 1.63 & & 0.29 &  0.29 & 0.975 &  \nodata & 1.19 \\
        Q1225$+$3145  & 2.016 & 62 & 0.30 & -0.07 & 2.03 & & 0.29 & -0.08 & 1.000 &  \nodata & 2.01 \\
        Q0421$+$019   & 1.870 & 70 & 0.54 &  0.72 & 1.60 & & 0.48 &  0.81 & 1.008 &  \nodata & 1.58 \\
        Q1225$+$3145  & 1.857 & 63 & 0.19 & -0.27 & 1.50 & & 0.17 & -0.60 & 0.995 &  \nodata & 1.39 \\
        Q0119$-$0437  & 1.807 & 83 & 0.34 &  0.80 & 1.32 & & 0.33 &  0.84 & 1.003 &  \nodata & 1.36 \\
        Q0058$+$0155  & 1.797 & 79 & 0.33 &  0.78 & 1.07 & & 0.31 &  0.87 & 1.010 &  \nodata & 1.01
   \enddata
   \tablenotetext{a}{Mean absorption redshift.}
   \tablenotetext{b}{Number of flux bins over which fit was performed.}
   \tablenotetext{c}{\hi\ ionization rate, in units of $10^{-1}$~s$^{-1}$.}
   \tablenotetext{d}{Power-law index for the generalized temperature-density
     relation $T^{0.7}\Gamma \propto \Delta^{\alpha}$.}
   \tablenotetext{e}{Factor by which to multiply the continuum
     in order for the model to produce the best fit.}
   \tablenotetext{f}{Flux zero point that would allow the model to
     produce the best fit.}
   \label{tab:mhr00_params_nonisothermal}
\end{deluxetable}

\clearpage

\clearpage

\clearpage

\clearpage

\clearpage

\clearpage

\clearpage

\clearpage

\clearpage

\clearpage

\clearpage

\clearpage

\clearpage

\clearpage

\clearpage

\begin{thebibliography}{}

\bibitem[Abel et al.(1997)]{abel97} Abel, T., Anninos, P., Zhang, Y.,
  \& Norman, M.~L.\ 1997, New Astronomy, 2, 181
\bibitem[Becker et al.(2006)]{becker06} Becker, G.~D., Sargent,
   W.~L.~W., Rauch, M., \& Simcoe, R.~A.\ 2006, \apj, 640, 69
\bibitem[Becker et al.(2001)]{becker01} Becker, R.~H., et al.\ 2001,
  \aj, 122, 2850
\bibitem[Bi et al.(1992)]{bi92} Bi, H.~G., Boerner, G., \& Chu, Y.\
   1992, \aap, 266, 1
\bibitem[Bi et al.(1995)]{bi95} Bi, H., Ge, J., \&
   Fang, L.-Z.\ 1995, \apj, 452, 90
\bibitem[Bi \& Davidsen(1997)]{bi97} Bi, H., \&
   Davidsen, A.~F.\ 1997, \apj, 479, 523
\bibitem[Bolton et al.(2004)]{bolton04} Bolton, J., Meiksin, A., \&
   White, M.\ 2004, \mnras, 348, L43
\bibitem[Choudhury et al.(2001)]{choudhury01} Choudhury, T.~R.,
   Srianand, R., \& Padmanabhan, T.\ 2001, \apj, 559, 29
\bibitem[Coles \& Jones(1991)]{coles91} Coles, P., \& Jones, B.\
   1991, \mnras, 248, 1
\bibitem[Desjacques \& Nusser(2005)]{desjacques05} Desjacques, V., \&
   Nusser, A.\ 2005, \mnras, 361, 1257
\bibitem[Fan et al.(2002)]{fan02} Fan, X., Narayanan, V.~K., Strauss,
  M.~A., White, R.~L., Becker, R.~H., Pentericci, L., \& Rix, H.-W.\
  2002, \aj, 123, 1247
\bibitem[Fan et al.(2006)]{fan06} Fan, X., et al.\ 2006, \aj, 132,
   117
\bibitem[Furlanetto et al.(2004)]{furlanetto04} Furlanetto, S.~R.,
  Hernquist, L., \& Zaldarriaga, M.\ 2004, MNRAS, 354, 695
\bibitem[Furlanetto et al.(2006)]{furlanetto06} Furlanetto, S.~R.,
  Zaldarriaga, M., \& Hernquist, L.\ 2006, \mnras, 365, 1012
\bibitem[Gazta{\~n}aga \& Croft(1999)]{gaztanaga99} Gazta{\~n}aga, E.,
  \& Croft, R.~A.~C.\ 1999, \mnras, 309, 885
\bibitem[Gunn \& Peterson(1965)]{gp65} Gunn, J.~E., \& Peterson,
  B.~A.\ 1965, \apj, 142, 1633
\bibitem[Haiman \& Cen(2005)]{haiman05} Haiman, Z., \& Cen, R.\ 2005,
  \apj, 623, 627
\bibitem[Hu \& Cowie(2006)]{hu06} Hu, E.~M., \& Cowie,
  L.~L.\ 2006, \nat, 440, 1145
\bibitem[Hu et al.(2004)]{hu04} Hu, E.~M., Cowie, L.~L., Capak, P.,
  McMahon, R.~G., Hayashino, T., \& Komiyama, Y.\ 2004, \aj, 127, 563
\bibitem[Hui \& Gnedin(1997)]{hui97} Hui, L., \& Gnedin, N.~Y.\ 1997,
  \mnras, 292, 27
\bibitem[Hui \& Haiman(2003)]{hui03} Hui, L., \& Haiman, Z.\ 2003,
  \apj, 596, 9
\bibitem[Kelson(2003)]{kelson03} Kelson, D.~D.\ 2003, \pasp, 115, 688
\bibitem[Kirkman et al.(2005)]{kirkman05} Kirkman, D., et al.\ 2005,
  \mnras, 360, 1373
\bibitem[Lidz et al.(2006a)]{lidz06a} Lidz, A., Oh, S.~P.,
  \& Furlanetto, S.~R.\ 2006a, \apjl, 639, L47
\bibitem[Lidz et al.(2006b)]{lidz06b} Lidz, A., Heitmann, K., Hui, L.,
  Habib, S., Rauch, M., \& Sargent, W.~L.~W.\ 2006b, \apj, 638, 27
\bibitem[Liu et al.(2006)]{liu06} Liu, J., Bi, H., Feng L.-L., \&
  Fang, L.-Z.\ 2006, ApJL, accepted (astro-ph/0605614)
\bibitem[Malhotra \& Rhoads(2006)]{mr06} Malhotra, S., \&
  Rhoads, J.~E.\ 2006, ApJL, submitted (astro-ph/0511196)
\bibitem[Malhotra \& Rhoads(2004)]{mr04} Malhotra, S., \&
  Rhoads, J.~E.\ 2004, ApJL, 617, L5
\bibitem[McDonald \& Miralda-Escud{\'e}(2001)]{mcdonald01} McDonald,
  P., \& Miralda-Escud{\'e}, J.\ 2001, \apjl, 549, L11
\bibitem[McDonald et al.(2000)]{mcdonald00} McDonald, P.,
  Miralda-Escud{\'e}, J., Rauch, M., Sargent, W.~L.~W., Barlow, T.~A.,
  Cen, R., \& Ostriker, J.~P.\ 2000, \apj, 543, 1
\bibitem[Mesinger et al.(2004)]{mesinger04b} Mesinger, A.,
  Haiman, Z., \& Cen, R.\ 2004, \apj, 613, 23
\bibitem[Mesinger \& Haiman(2004)]{mesinger04a} Mesinger, A.,
  \& Haiman, Z.\ 2004, \apjl, 611, L69
\bibitem[Miralda-Escud{\'e} et al.(2000)]{me00} Miralda-Escud{\'e},
  J., Haehnelt, M., \& Rees, M.~J.\ 2000, \apj, 530, 1
\bibitem[Miralda-Escud{\'e} et al.(1996)]{me96} Miralda-Escud{\'e},
  J., Cen, R., Ostriker, J.~P., \& Rauch, M.\ 1996, \apj, 471, 582
\bibitem[Oh \& Furlanetto(2005)]{ohfur05} Oh, S.~P., \& Furlanetto,
  S.~R.\ 2005, \apjl, 620, L9
\bibitem[Press et al.(1992)]{press92} Press, W.~H., Teukolsky, S.~A.,
  Vetterling, W.~T., \& Flannery, B.~P.\ 1992, Numerical Recipers in C
  (2nd Ed.; Cambridge: Cambridge Univ.  Press)
\bibitem[Rauch et al.(1997)]{rauch97} Rauch, M., et al.\ 1997, \apj,
  489, 7
\bibitem[Rauch(1998)]{rauch98} Rauch, M.\ 1998, \araa, 36, 267
\bibitem[Santos(2004)]{santos04} Santos, M.~R.\ 2004, MNRAS, 349, 1137
\bibitem[Songaila \& Cowie(2002)]{songaila02} Songaila, A., \& Cowie,
  L.~L.\ 2002, \aj, 123, 2183
\bibitem[Songaila(2004)]{songaila04} Songaila, A.\ 2004, \aj, 127,
  2598
\bibitem[Stern et al.(2005)]{stern05} Stern, D., Yost, S.~A., Eckart,
  M.~E., Harrison, F.~A., Helfand, D.~J., Djorgovski, S.~G., Malhotra,
  S., \& Rhoads, J.~E.\ 2005, ApJ, 619, 12
\bibitem[Suzuki(2006)]{suzuki06} Suzuki, N.\ 2006, \apjs, 163, 110
\bibitem[Telfer et al.(2002)]{telfer02} Telfer, R.~C.,
  Zheng, W., Kriss, G.~A., \& Davidsen, A.~F.\ 2002, \apj, 565, 773
\bibitem[Theuns et al.(2002)]{theuns02} Theuns, T., Schaye, J.,
  Zaroubi, S., Kim, T.-S., Tzanavaris, P., \& Carswell, B.\ 2002,
  \apjl, 567, L103
\bibitem[Vogt et al.(1994)]{vogt94} Vogt, S.~S., et al.\ 1994,
  \procspie, 2198, 362
\bibitem[Weinberg et al.(1997)]{weinberg97} Weinberg, D.~H.,
  Miralda-Escud{\'e}, J., Hernquist, L., \& Katz, N.\ 1997, \apj, 490, 564
\bibitem[White et al.(2003)]{white03} White, R.~L., Becker, R.~H.,
  Fan, X., \& Strauss, M.~A.\ 2003, \aj, 126, 1
\bibitem[White et al.(2005)]{white05} White, R.~L., Becker, R.~H.,
  Fan, X., \& Strauss, M.~A.\ 2005, \aj, 129, 2102
\bibitem[Wyithe \& Loeb(2004)]{wyithe04} Wyithe, J.~S.~B.,
  \& Loeb, A.\ 2004, \nat, 427, 815
\bibitem[Wyithe \& Loeb(2005)]{wyithe05} Wyithe, J.~S.~B.,
  \& Loeb, A.\ 2005, \apj, 625, 1

\end{thebibliography}
\end{document}